\documentclass[10pt,letterpaper,amsmath,amssymb,preprint,showpacs]{elsarticle}
\usepackage{longtable}
\usepackage{hyperref}
\usepackage{pdflscape}
\usepackage{caption}
\usepackage{graphicx}
\usepackage{dcolumn}
\usepackage{multirow}
\usepackage{bbm}
\usepackage{textcomp}
\usepackage{color}
\usepackage{gensymb}
\usepackage{wasysym}
\usepackage{amssymb}
\usepackage{rotating}

\newcounter{rownumber}
\def\nuc#1#2{${}^{#1}$#2}

\def\BBz{$\beta\beta(0\nu)$}

\def\be{\begin{equation}}
\def\ee{\end{equation}}
\def\cpRty{c/(ROI t yr)}
\def\onecpRty{1~c/(ROI t yr)}
\def\threecpRty{3~c/(ROI t yr)}

\def\MJ{{\sc Majorana}}             
\def\DEM{{\sc Demonstrator}}             
\def\mj{{\sc Majorana}}             
\def\dem{{\sc Demonstrator}}             

\def\Rmark{\textsuperscript{\textregistered}}
\def\Tmark{\texttrademark}

\newcommand{\mcc}[2]{\multicolumn{#1}{c}{#2}}
\newcommand{\mcl}[2]{\multicolumn{#1}{l}{#2}}

\journal{Nuclear Inst. and Methods in Physics Research, A}

\begin{document}

\begin{frontmatter}

\title{The M{\sc ajo\-ra\-na} D{\sc e\-mon\-strat\-or} Radioassay Program}

\author[lbnl]{N.~Abgrall}		
\author[pnnl]{I.J.~Arnquist}
\author[usc,ornl]{F.T.~Avignone~III}
\author[ncsu,tunl]{H.O.~Back\fnref{Henning}}
\author[ITEP]{A.S.~Barabash}	
\author[ornl]{F.E.~Bertrand}
\author[lanl]{M.~Boswell} 
\author[lbnl]{A.W.~Bradley}
\author[JINR]{V.~Brudanin}
\author[duke,tunl]{M.~Busch}	
\author[uw]{M.~Buuck}
\author[usd]{D.~Byram} 
\author[sdsmt]{A.S.~Caldwell}
\author[lbnl]{Y-D.~Chan}
\author[sdsmt]{C.D.~Christofferson} 
\author[lanl]{P.-H.~Chu}
\author[uw]{C.~Cuesta}	
\author[uw]{J.A.~Detwiler}	
\author[uw]{J.A.~Dunmore}
\author[ut]{Yu.~Efremenko}
\author[ou]{H.~Ejiri}
\author[lanl]{S.R.~Elliott\corref{cor1}}
\author[unc,tunl]{P.~Finnerty\fnref{Paddy}}
\author[ornl]{A.~Galindo-Uribarri}	
\author[lanl]{V.M.~Gehman\fnref{Vic}}
\author[unc,tunl]{T.~Gilliss}
\author[unc,tunl]{G.K.~Giovanetti}  
\author[lanl]{J. Goett}	
\author[ornl,ncsu,tunl]{M.P.~Green}  
\author[uw]{J. Gruszko}		
\author[uw]{I.S.~Guinn}
\author[usc]{V.E.~Guiseppe}	
\author[unc,tunl]{R.~Henning}
\author[pnnl]{E.W.~Hoppe}
\author[sdsmt]{S. Howard}  
\author[unc,tunl]{M.A.~Howe}
\author[usd]{B.R.~Jasinski}
\author[uw]{R.A.~Johnson\fnref{Rob}}
\author[blhill]{K.J.~Keeter}
\author[ttu]{M.F.~Kidd}	
\author[JINR]{O.~Kochetov}
\author[ITEP]{S.I.~Konovalov}
\author[pnnl]{R.T.~Kouzes}
\author[pnnl]{B.D.~LaFerriere}   
\author[uw]{J.~Leon}	
\author[sjtu]{J.C.~Loach\fnref{James}}	
\author[unc,tunl]{J.~MacMullin}
\author[unc,tunl]{S.~MacMullin\fnref{Sean}}
\author[usd]{R.D.~Martin\fnref{Ryan}}
\author[lanl]{R. Massarczyk}
\author[unc,tunl]{S. Meijer}	
\author[lbnl]{S.~Mertens}		
\author[uw]{M.L.~Miller}
\author[pnnl]{J.L.~Orrell}
\author[unc,tunl]{C.~O'Shaughnessy}	
\author[pnnl]{N.R.~Overman}  
\author[lbnl]{A.W.P.~Poon}
\author[usd]{K.~Pushkin\fnref{Kirill}}
\author[ornl]{D.C.~Radford}
\author[unc,tunl]{J.~Rager}	
\author[lanl]{K.~Rielage}
\author[uw]{R.G.H.~Robertson}
\author[ut,ornl]{E.~Romero-Romero} 
\author[lanl]{M.C.~Ronquest\fnref{MikeR}}
\author[uw]{A.G.~Schubert\fnref{Alexis}}
\author[unc,tunl]{B.~Shanks}	
\author[JINR]{M.~Shirchenko}
\author[unc,tunl]{K.J.Snavely\fnref{Kyle}}
\author[usd]{N.~Snyder}	
\author[lanl]{D.~Steele\fnref{David}}
\author[sdsmt]{A.M.~Suriano} 
\author[usc]{D.~Tedeschi}
\author[unc,tunl]{J.E.~Trimble}
\author[ornl]{R.L.~Varner}  
\author[ut]{S.~Vasilyev}
\author[lbnl]{K.~Vetter\fnref{ucb}}
\author[unc,tunl]{K.~Vorren} 
\author[ornl]{B.R.~White}	
\author[unc,tunl,ornl]{J.F.~Wilkerson}    
\author[usc]{C.~Wiseman}		
\author[lanl]{W.~Xu}  
\author[JINR]{E.~Yakushev}
\author[ornl]{C.-H.~Yu}
\author[ITEP]{V.~Yumatov}
\author[JINR]{I.~Zhitnikov}

\cortext[cor1]{Corresponding author, elliotts@lanl.gov}
\address[lbnl]{Nuclear Science Division, Lawrence Berkeley National Laboratory, Berkeley, CA, USA}
\address[pnnl]{Pacific Northwest National Laboratory, Richland, WA, USA}
\address[usc]{Department of Physics and Astronomy, University of South Carolina, Columbia, SC, USA}
\address[ornl]{Oak Ridge National Laboratory, Oak Ridge, TN, USA}
\address[ncsu]{Department of Physics, North Carolina State University, Raleigh, NC, USA}
\address[tunl]{Triangle Universities Nuclear Laboratory, Durham, NC, USA}
\address[ITEP]{National Research Center ``Kurchatov Institute'' Institute for Theoretical and Experimental Physics, Moscow, Russia}
\address[lanl]{Los Alamos National Laboratory, Los Alamos, NM, USA}
\address[JINR]{Joint Institute for Nuclear Research, Dubna, Russia}
\address[duke]{Department of Physics, Duke University, Durham, NC, USA}
\address[usd]{Department of Physics, University of South Dakota, Vermillion, SD, USA} 
\address[sdsmt]{South Dakota School of Mines and Technology, Rapid City, SD, USA}
\address[uw]{Center for Experimental Nuclear Physics and Astrophysics, and Department of Physics, University of Washington, Seattle, WA, USA}
\address[ut]{Department of Physics and Astronomy, University of Tennessee, Knoxville, TN, USA}
\address[ou]{Research Center for Nuclear Physics and Department of Physics, Osaka University, Ibaraki, Osaka, Japan}
\address[unc]{Department of Physics and Astronomy, University of North Carolina, Chapel Hill, NC, USA}
\address[alberta]{Centre for Particle Physics, University of Alberta, Edmonton, AB, Canada}
\address[blhill]{Department of Physics, Black Hills State University, Spearfish, SD, USA} 
\address[ttu]{Tennessee Tech University, Cookeville, TN, USA}
\address[sjtu]{Shanghai Jiao Tong University, Shanghai, China}
\fntext[Henning]{Permanent Address: Pacific Northwest National Laboratory, Richland, WA, USA}
\fntext[Paddy]{Permanent Address: Applied Research Associates, Inc., 8537 Six Forks Road, Suite 600, raleigh, NC, USA}
\fntext[Rob]{Permanent Address: Microsoft Corporation, One Microsoft Way, Redmond, WA, USA}
\fntext[James]{Permanent Address: Shanghai Jiao Tong University, Shanghai, China}
\fntext[Sean]{Permanent Address:  Picarro Inc., 3105 Patrick Henry Dr., Santa Clara, CA, USA}
\fntext[Vic]{Permanent Address: Lawrence Berkeley National Laboratory, Berkeley, CA, USA}
\fntext[Kirill]{Permanent Address: Randall Laboratory of Physics, University of Michigan, Ann Arbor, Michigan, USA}
\fntext[MikeR]{Permanent Address: CCRi, 1422 Sachem Pl \#1, Charlottesville, VA, USA}
\fntext[Alexis]{Permanent Address: Department of Physics, Stanford University, Stanford, CA, USA}
\fntext[David]{Permanent Address: Picarro Inc., 3105 Patrick Henry Dr., Santa Clara, CA, USA}
\fntext[Kyle]{Permanent Address: IBM Cloudant, Boston, 200 State St, Boston, MA, USA}
\fntext[Ryan]{Permanent Address: Department of Physics, Engineering Physics and Astronomy, Queen's University, Kingston, ON, Canada}  
\fntext[ucb]{Alternate Address: Department of Nuclear Engineering, University of California, Berkeley, CA, USA}


\date{\today}

\begin{abstract}
The \mj\ collaboration is constructing the \mj\ \dem\ at the Sanford Underground Research Facility at the Homestake gold mine, in Lead, SD. The apparatus will use Ge detectors, enriched in isotope \nuc{76}{Ge}, to demonstrate the feasibility of a large-scale Ge detector experiment to search for neutrinoless double beta decay. The long half-life of this postulated process requires that the apparatus be extremely low in radioactive isotopes whose decays may produce backgrounds to the search. The radioassay program conducted by the collaboration to ensure that the materials comprising the apparatus are sufficiently pure is described. The resulting measurements from gamma-ray counting, neutron activation and mass spectroscopy of the radioactive-isotope contamination for the materials studied for use in the detector are reported. We interpret these numbers in the context of the expected background for the experiment. 
\end{abstract}



\end{frontmatter}

\section{Introduction: Overview of the \dem}
\begin{figure}[htbp]
  \centering
      \includegraphics[width=0.75\textwidth]{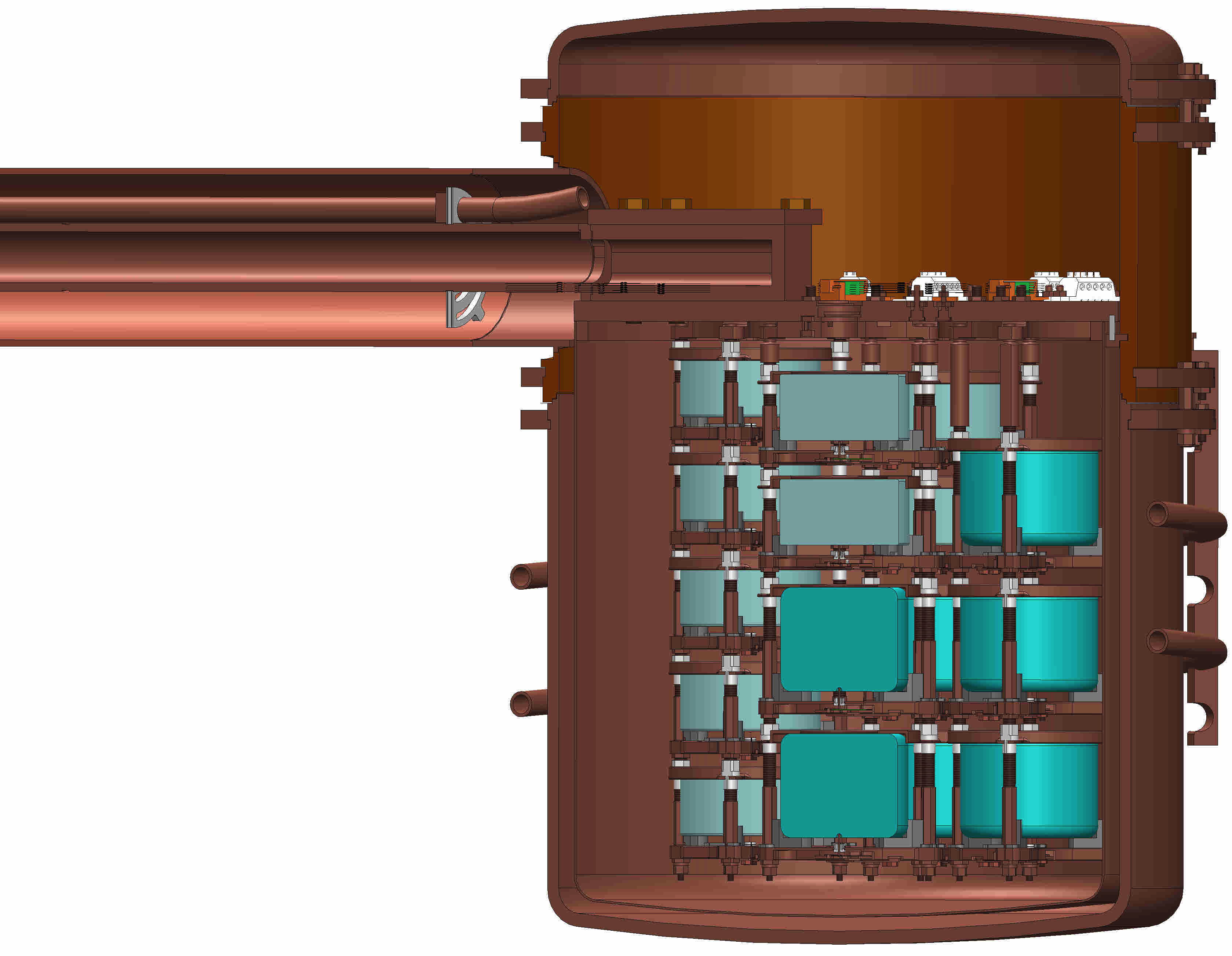}
\caption{A cross sectional view of a \mj\ \dem\ cryostat.  The strings within each cryostat hold a mixture of natural and enriched germanium detectors.\label{fig:cryo_internals}} 
\end{figure}

\begin{figure}[htbp]
  \centering
      \includegraphics[width=0.75\textwidth]{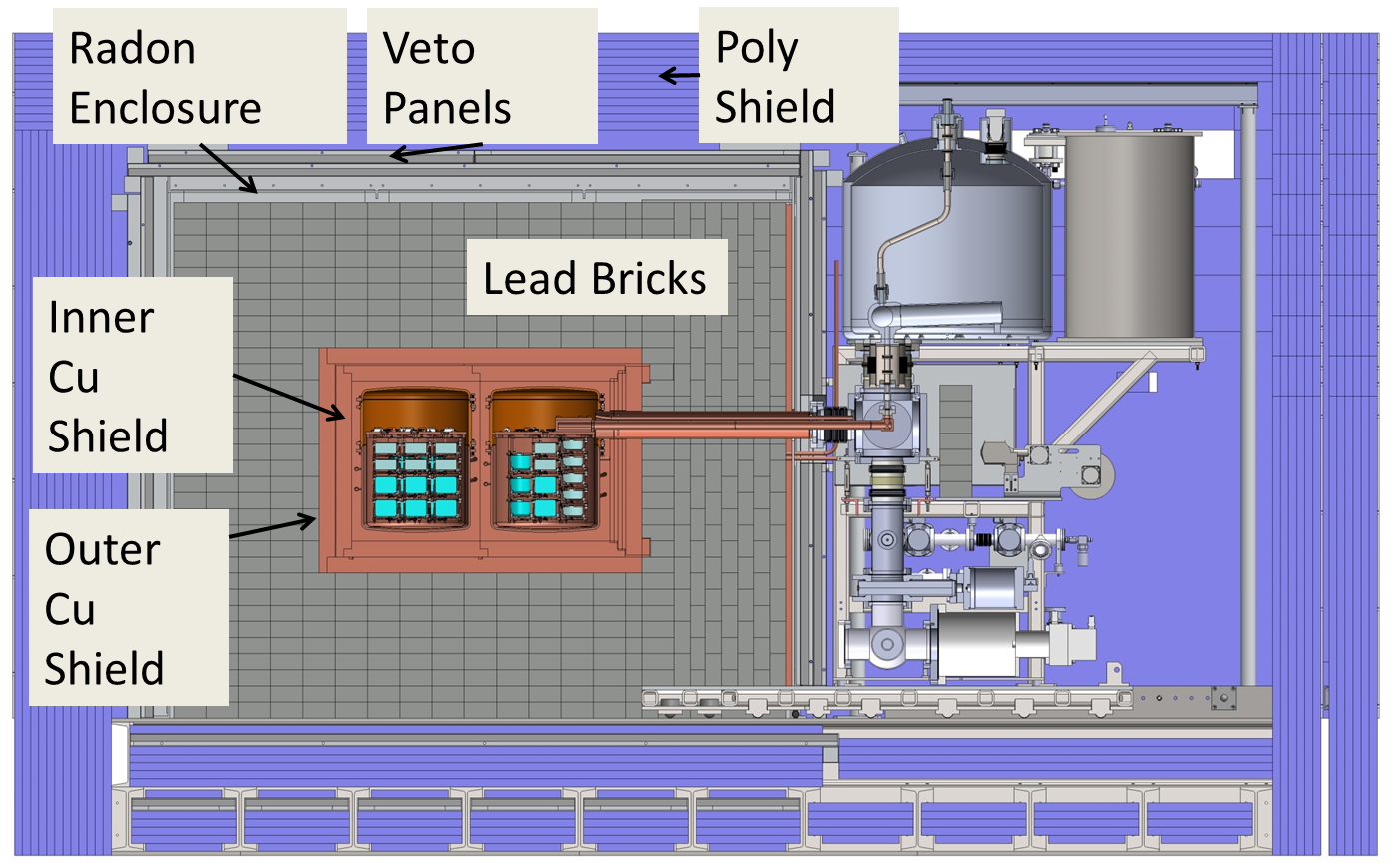}
\caption{The \mj\ \dem\ is shown here with both active and passive shielding in place.  The outer surface of the inner Cu shield is 50.8 cm in height and 76.2 cm in length. \label{fig:mjd}} 
\end{figure}

The \mj\ collaboration~\cite{Abgrall2014} will search for the neutrinoless double-beta decay (\BBz) of \nuc{76}{Ge}. The observation of this rare decay would indicate the neutrino is its own anti-particle, demonstrate that lepton number is not conserved, and provide information on the absolute mass-scale of the neutrino (see Refs.~\cite{ell02,ell04,Avi08,Rode11,Bar11,GomezCadenas11,Vergados2012} for recent reviews of \BBz). Reaching the neutrino mass-scale sensitivity associated with the inverted mass ordering ($15 - 50$ meV) is a goal for next-generation \BBz\ searches. This goal will require a half-life sensitivity exceeding 10$^{27}$~y, which corresponds to a signal on the order of a few counts or fewer per tonne-year in the \BBz\ peak (2039 keV for \nuc{76}{Ge}). To observe such a rare signal, one will need to construct large-scale experiments with backgrounds in the region of interest (ROI) below 1 count per tonne of isotope per year ($<$\onecpRty).  The \mj\ collaboration~\cite{Abgrall2014} is constructing the \dem, an array of high-purity germanium (HPGe) detectors at the 4850 ft level of the Sanford Underground Research Facility (SURF) in Lead, South Dakota~\cite{Heise2014,Heise2015}. The \dem\ will consist of a mixture of HPGe detectors including, 15 kg fabricated from natural-isotopic-abundance Ge and 29.7 kg fabricated from Ge enriched to $>$87\% in \nuc{76}{Ge}. These detectors are contained within two low-background copper cryostats. Each cryostat will contain seven closely-packed stacks of detectors, called strings, with up to five detectors comprising each string. Figures~\ref{fig:cryo_internals}~and~\ref{fig:mjd} show the \dem\ concept.  

The \dem\ aims to show that a background rate lower than \threecpRty\  in the 4 keV ROI surrounding the 2039 keV \nuc{76}{Ge}\ Q-value energy is achievable. This background level will scale to \onecpRty\ in a future experiment based on simulations considering improved self-shielding, thicker inner copper shield, and improved cosmogenic isotope control. Hence the \dem\ will establish the technology required to build a large-scale germanium based \BBz\ experiment.

The \mj\ collaboration uses p-type point contact (PPC) HPGe detectors.  These detectors \cite{luk89,Barbeau2007,Aguayo2013,Xu2014} have been demonstrated to provide both good energy resolution ($<$2.0 keV FWHM at 1332 keV, $<$4.0 keV FWHM at 2039 keV) and low-energy threshold ($\sim$500 eV) \cite{Barbeau2007,Aal08}.  (See e.g., Fig.~1 in Ref.~\cite{Aal08}.)  
Each PPC detector used in the \dem\  has a mass in the range of 0.5-1.1~kg with a mean of 0.85 kg.

This report summarizes the assay program conducted by the \mj\ collaboration to ensure that the various components of the experimental apparatus have radioactive isotope contamination sufficiently low to meet the background goal.  Section~\ref{Sec:LBGStrategy}  discusses the strategy to reach the required background levels in the experiment. Section~\ref{Sec:Methods}  describes the methods and facilities used for the assays. Section~\ref{Sec:Results}  summarizes the results giving the levels of radioactive contamination found in the various materials studied and describes some special techniques or materials that were investigated. Finally, Section~\ref{Sec:Conclusions}  provides some conclusions based on these studies.

Numerous other such studies have been done and, in concert with the results here, there is a wealth of information available to help select materials for future projects. Other reports can be found in Refs.~\cite{Rau2000,Apresella2002,Leonard2008,Aprile2011,Armengaud2013,Ahmed2014,Akerib2015,ILIASdb,GOPHER}. Heusser~\cite{Heusser1995} wrote a nice review on low background counting techniques. The results presented in this manuscript will be made available on the online database at radiopurity.org following publication.

\section{The \dem\ Low-Background Strategy and Model}\label{Sec:LBGStrategy}
The projected background in the \DEM\ is significantly lower than previous generation experiments. This reduction is achieved in part by fielding the detectors in large arrays that share a cryostat, allowing for the minimization of the amount of interstitial passive material. Further background suppression is achieved through the aggressive reduction of radioactive impurities in construction materials and by minimizing exposure to cosmic rays.  \MJ\ will also make use of event signatures including pulse-shape characteristics, detector hit granularity, cosmic ray veto tags, and single-site time correlations to reduce backgrounds that do appear. In this section we describe these aspects of the \MJ\ \DEM\ design and their impact on the projected backgrounds and physics sensitivity. In future sections we add detail about processes that are not published elsewhere and cite those that are.

The production process for enriched germanium detectors (enrichment, zone refining, and crystal growth) efficiently removes natural radioactive impurities from the bulk germanium. The cosmogenic activation isotopes, \nuc{60}{Co} and \nuc{68}{Ge}, are produced in the crystals while they are above ground, but can be sufficiently reduced by limiting the time  above ground and by the use of passive shielding during transport and storage. 

For the main structural material in the innermost region of the apparatus, \MJ\ chose copper for its lack of naturally occurring radioactive isotopes and its excellent physical properties. Starting with the cleanest copper stock identified and electroforming it underground reduces primordial radioactivity and cosmogenically-produced \nuc{60}{Co}. This electroformed copper will also be employed for the innermost passive, high-Z shield. Commercial oxygen-free high-conductivity (OFHC) copper stock is clean enough for use as the next layer of shielding. For all uses of copper, \MJ\  certified the cleanliness of samples via assay. Modern lead is available with sufficient purity for use as the bulk shielding material outside of the copper layers.

Several clean plastics are available for electrical and thermal insulation. For the detector supports we use a pure polytetrafluoroethylene (PTFE), DuPont\Tmark~\\Teflon\Rmark~NXT-85.  For the few weight-bearing plastic components requiring higher rigidity, we have sourced pure stocks of PEEK\Rmark~(polyether ether ketone), produced by Victrex\Rmark, and Vespel\Rmark (all grade SP-1), produced by DuPont\Tmark. After machining, these parts are leached to remove surface activity. Thin layers of low-radioactivity parylene are used as a coating on copper threads to prevent galling, and for providing electrical insulation.

The high material purities required for the \MJ\ \DEM\ necessitated the development of improved assay capabilities. These capabilities are needed not only to ensure that the required purities can be achieved, but to also monitor construction processes to verify that cleanliness is maintained. We rely primarily on three assay methods: $\gamma$-ray counting,  mass spectrometry, and neutron activation analysis. 

Unlike activation analysis and mass spectrometry, $\gamma$-ray counting has the benefit of being non-destructive so that specific assayed parts can be used for the apparatus, hence avoiding a reliance on sampling. However, it is not sensitive enough for most parts of the \DEM\ and is used primarily for prescreening and for parts requiring only modest radiopurity. Furthermore, it is also useful to supplement mass spectrometry with $\gamma$-ray counting to cross-check for potential disequilibria in the natural decay chains. Mass spectrometry and neutron activation analysis are destructive methods that only measure the concentrations of the isotopes \nuc{238}{U} and \nuc{232}{Th}, which can elucidate the concentrations of  \nuc{214}{Bi} and \nuc{208}{Tl} when secular equilibrium can be assumed. However, \nuc{238}{U} and \nuc{232}{Th} are at the top of these chains, unlike \nuc{214}{Bi} and \nuc{208}{Tl}, which produce the $\gamma$ rays that actually comprise the radioactive background. By contrast, $\gamma$-ray counting measures the \nuc{214}{Bi} and \nuc{208}{Tl} activities directly.

\section{Description of Assay Methods and Facilities}\label{Sec:Methods}
\subsection{$\gamma$-Ray Counting}
\label{Sec:GammaCount}

For $\gamma$-ray counting, the \mj\ collaboration primarily used three facilities. Each had its own Monte Carlo for estimating efficiencies for each sample counted. A well-characterized sample was counted on all systems and analyzed blind as the operators did not know the activity of this test sample. All groups found similar results to within about 20\%. The uncertainties in the results were dominated by counting statistics and simulation details. All the detectors were enclosed within a Cu shield surrounded by a Pb brick enclosure. All of the cavities were purged of Rn with boil-off gas from liquid nitrogen.

The three facilities have their key characteristics summarized in Table~\ref{tab:GammaCount} and were:

\begin{itemize}
\item The low-background counting facility located underground at the Waste Isolation Pilot Plant (WIPP) near Carlsbad, NM operated by Los Alamos National Laboratory. (See, e.g., Ref.~\cite{Ell10}.) The Ge detector used at this facility was fabricated in 1985 and placed underground at WIPP in 1998. It is an n-type semi-coax design.  This facility was decommissioned in 2014.

\item The Kimballton Underground Research Facility (KURF)~\cite{Finnerty2011} is located at Lhoist North America's Kimballton mine in Ripplemead, VA and is operated by the Virginia Polytechnic Institute and State University. The counting facility, operated by the University of North Carolina and Triangle Universities Nuclear Laboratory, consists of two HPGe detectors specifically designed for low-background assay work. The first detector, named VT-1, is a commercial ORTEC LLB (very low-background) series coaxial detector. The second detector, named MELISSA, is a Canberra LB (low background) coaxial detector.  

\item The low background facility (LBF) at Oroville~\cite{Thomas2013,Thomas2015}, operated by Lawrence Berkeley National Laboratory (LBNL), was located in the Edward Hyatt Power Plant of the Oroville Dam in Oroville, CA.  The p-type HPGe/ULB detector  (manufactured by ORTEC) has been underground since $\sim$1995. This apparatus was moved to SURF in 2014. A separate low background facility in Building 72 of LBNL was used to count samples that do not require high sensitivity and to pre-screen samples for more sensitive counting at other facilities.  The principal detector is an n-type germanium detector.
\end{itemize}

\begin{table}[]
\centering
\caption{A summary of the key characteristics of the 5 Ge detectors housed at 3 facilities that were used for this work. The overburden is given in meters water equivalent (mwe). All the cavity footprints are square except for VT-1, which is circular.}
\label{tab:GammaCount}
\begin{tabular}{llllll}
\hline
                                                                         & LBNL & LBNL & KURF & KURF & WIPP                            \\
                                                                         &Surface& Oroville & MELISSA &  VT-1 &                             \\
 \hline\hline
Overburden                                                               & 10 mwe                                              & 600 mwe                                             & 1450 mwe                                            & mwe                                        & 1700 mwe                     \\
Rel. Effic.                     & 115\%                                                  & 85\%                                                    & 50\%                                                   & 35\%                                                & 22\%                            \\
Shield Cu                &                                                        & 12.7 mm                                                 & 25.4 mm                                                & 0.3 mm                                              & 50 mm                           \\
Cavity Height               &      14.5~cm                                                  & 43~cm                        & 38~cm                       & 41~cm            & 15~cm \\
Cavity Area                   &   317~cm$^2$                     & 400~cm$^2$                         & 1444~cm$^2$                       & 616~cm$^2$           & 100~cm$^2$ \\
$\approx$ U Sens.       & 0.5 ppb                                                & 50 ppt                                                  &   0.4 ppb                                                     &     0.7 ppb                                             & 0.1 ppb                         \\
$\approx$ Th Sens.      & 2.0 ppb                                                & 200 ppt                                                 &   0.8 ppb                                                     &    1.6 ppb                                                 & 0.3 ppb                         \\
\hline
\end{tabular}
\end{table}

\subsection{Mass Spectrometry Analysis}
\label{Sec:ICP-MS}
Both Glow Discharge Mass Spectrometry (GDMS) and Inductively-Coupled Plasma Mass Spectrometry (ICPMS) were used for \MJ\ \DEM\ assays. Each of these measure specific elements within the decay chains and therefore give little information about the equilibrium of the chain. Our background estimates assume equilibrium and therefore, since our most sensitive assays come from ICPMS, there is an uncertainty associated with this assumption that is difficult to estimate.

\subsubsection{Glow Discharge Mass Spectrometry}
\label{Sec:GSMS}

Glow Discharge Mass Spectrometry is useful for electrically conductive samples and requires very little sample preparation since the surface can be sputtered cleanly.  Sensitivities approaching ppt ($10^{-12}$~g/g) can be achieved for high mass elements such as U and Th. A homogenous sample is required unless surface-only or depth profiling is desired.  In GDMS the bombarding ions come from a low pressure DC plasma discharge cell in which the sample is the cathode. GDMS is not particularly matrix-dependent and can be performed directly on samples with little or no preparation or separation chemistry. It also offers the advantage of quick turnaround compared to other mass spectrometric analysis methods that require time- and labor-intensive preparation.  Conductive materials are easiest to analyze by GDMS, although conductive electrodes can often be formed when the sample material is non-conductive.  GDMS is excellent for identifying trace elements in bulk samples down to tens of ppt. \MJ\ used the Thermo VG9000 GDMS instrument at the National Research Council of Canada (NRCC)~\cite{NRCC} for Pb assay and an Astrum at Nu Instruments Limited in Wrexham, UK~\cite{NUinst} to provide K values for the electroformed copper. 

\subsubsection{Inductively-Coupled Plasma Mass Spectrometry }
\label{Sec:ICPMS}
A sample analyzed by Inductively-Coupled Plasma Mass Spectrometry  is usually first put into solution using various combinations of acids or bases.  In ICPMS, a flow of gas (usually argon) converts the liquid sample into a fine aerosol. A portion of the aerosol is then directed through the center of an argon plasma torch, where the aerosol particles are effectively atomized and ionized. In some cases the ions are then directed into a hexapole or octopole collision cell where polyatomics (molecular ions) can be dissociated or excluded before entering the final mass filter, most often a quadrupole.   \MJ\ used four facilities for its ICPMS analysis. Most of our assay analyses were performed at Pacific Northwest National Laboratory (PNNL) and Lawrence Berkeley National Laboratory (LBNL); those facilities are described below. \MJ\ also used the Institute of Microelectronic Technology and High Purity Materials in Chernogolovka, Russia~\cite{IMTRAS} and Validation Resources Inc in Bend OR~\cite{VRAnal}. 

In general, ICPMS samples for introduction must be prepared in solution.  This requires solid samples to be digested prior to assay.  For the utmost in accuracy and precision, we use isotope tracer dilution methods to account for sample preparation effects on the analyte as well as plasma perturbations and instrument drift during analysis. While sample dissolution complicates the procedure, the advantage is that the results represent the bulk material which is homogenized into a solution. This also allows for solution-based chemistries to be performed, if desired, to further concentrate the analyte or remove unwanted matrices.  However, the sample size is typically small, so for large items only an indication of the bulk contamination results. For compound materials, some components dissolve better than others. For \MJ\ we demonstrated sensitivities better than ppt for U and Th using ICPMS.

Samples were prepared in clean room conditions to avoid contamination. Elemental isolation or chemical purification techniques were often necessary to avoid the isobaric interferences caused by other isotopes or ion complexes within large quantities of dissolved solids.  Based on the \MJ\ \DEM\  background calculations, the most stringent radiopurity goal is that for copper used in the inner shield and detector components. The effort has reached the required purity levels ($<$0.3 $\mu$Bq for both \nuc{232}{Th} and \nuc{238}{U} per kg of copper or $0.024 \times 10^{-12}$ g \nuc{238}{U}/g Cu and $0.075 \times 10^{-12}$ g \nuc{232}{Th}/g Cu). See Ref.~\cite{Hoppe2014,LaFerriere2015} and the results reported here in Rows~\ref{row:r1} through \ref{row:r203} in Table~\ref{tab:assay}. 

At LBNL the ICPMS measurements were done at the Earth Sciences Division's Aqueous Geochemistry Laboratory. This facility is equipped with a PerkinElmer SCIEX Elan DRCII ICPMS instrument and an Anton Paar Multiwave 3000 Microwave Reaction System, as well as Class 100 flow hoods for sample and standards preparation. This equipment has been used for quantitative ppt-scale measurements in a variety of materials.

At PNNL an Agilent 7700 ICPMS and all sample preparation equipment is located in dedicated clean rooms for low background measurements.  With a variety of advanced dissolution technologies, such as electrochemical sample preparation or microwave digestion, virtually anything can be brought into solution and analyzed by ICPMS down to the sub ppt or ppq level. All analysis performed at PNNL for the \MJ\ \DEM\ used isotopic tracers as the standard.  The tracers used for the analysis of \nuc{232}{Th} and \nuc{238}{U} were  \nuc{229}{Th} and  \nuc{233}{U}, respectively.

\subsection{Neutron Activation Analysis}
\label{Sec:NAA}
For materials such as hydrocarbons with no long-lived neutron
activation products, instrumental neutron activation analysis (NAA) can achieve substantially greater sensitivity than direct $\gamma$-ray counting. In this technique, samples are irradiated with neutrons from a nuclear reactor. When the neutrons are captured on \nuc{238}{U} and \nuc{232}{Th}, the isotopes \nuc{239}{Np} and \nuc{233}{Pa} are generated. After irradiation, the samples are counted by high-resolution $\gamma$-ray detectors to search for characteristic lines at 106 keV and 312 keV from \nuc{239}{Np} and \nuc{233}{Pa} decays. Using known or calibrated neutron capture probabilities, neutron flux, irradiation time, $\gamma$-ray detector efficiencies, and sample mass, the concentrations of U and Th in a sample can be calculated.

For high-sensitivity NAA it is important to have samples free from contaminants. Plastic samples were prepared for assay by leaching with Optima\Tmark\ grade (Fisher Scientific) acids. After leaching, samples were dried with nitrogen and handling was minimized thereafter to avoid the introduction of Na or K. These elements can easily capture neutrons producing the short-lived \nuc{24}{Na} and \nuc{42}{K} isotopes. The contribution to the $\gamma$-ray spectra from these elements can obscure lines from \nuc{239}{Np} and \nuc{233}{Pa} that indicate the presence of U and Th. Other contaminants that can decrease sensitivity and require care to avoid include Cr, Mn, Cu, Zn, Br, W, and Au.   

\MJ\ made use of three neutron activation facilities. These facilities have their key characteristics summarized in Table~\ref{Tab:NAAFacilities} and were:

\begin{itemize}
\item  The University of California Davis' McClellan Nuclear Radiation Center (MNRC) provides an array of options for sample irradiations including both in-core and out-of-core locations with varying encapsulation requirements, neutron spectra, and total neutron fluence limits. For this work we employed MNRC's Pneumatic Transfer System, in which small samples are placed inside polyethylene containers (rabbits) and transferred into and out of the reactor core via a pneumatic system. Consideration of the structural integrity of the rabbits limits irradiations for in-core position to $<$1 MW-hour.  Neutron fluences were calibrated by simultaneously irradiating a sample of ``standard pottery'' ceramic material, with well-characterized U and Th content, in the same rabbit as the sample of interest. Post-irradiation sample counting was performed at the low-background counting facilities at LBNL. (See Section~\ref{Sec:GammaCount}.)

\item North Carolina State University's research and teaching reactor, in its Nuclear Engineering Department, uses the 1 MW PULSTAR reactor facility for irradiations. Our samples were irradiated in an out-of-core port in which the sample was rotated to provide a smooth irradiation profile across the sample holder. Our samples were irradiated for 12 MW-hours along with samples of standard pottery and other flux-monitoring materials used by the NC State facility. Assay of the samples post-irradiation was performed on-site via HPGe $\gamma$-ray detectors.

\item The High Flux Isotope Reactor (HFIR) at Oak Ridge National Laboratory (ORNL) is an 85 MW enriched fuel reactor located at ORNL. Samples are irradiated at HFIR by pneumatically inserting rabbits into the beryllium reflector at a distance of 180 mm from the edge of the fuel region. The neutron flux at the irradiation location is $2.8 \times 10^{14}$ thermal n/(cm$^2$ s). The thermal-to-epithermal flux ratio is approximately 40. Sample size is limited by the size of the rabbit and is usually at the level of a few grams or less. Irradiation time is limited to 10-20 minutes to prevent sample melting due to extensive $\gamma$-ray heating. After irradiation, the samples were left overnight in a hot cell to allow highly-radioactive short-lived isotopes to decay and then counted on-site via HPGe $\gamma$-ray detectors. Some samples were also subsequently counted underground at KURF with a low background Ge detector. (See Section~\ref{Sec:GammaCount}.)

\end{itemize}

\begin{table}[]
\centering
\caption{A summary of the key characteristics of the NAA facilities used for this work.}
\label{Tab:NAAFacilities}
\begin{tabular}{llll}
\hline
Facility       & MNRC                      & NCSU                          & HFIR                            \\
\hline\hline
Reactor        & TRIGA                     & PULSTAR                       & Enriched Fuel                   \\
Power          & 2 MW                      & 1 MW                          & 85 MW                           \\
Thermal n Flux & $1\times10^{13}$~n/cm$^2$ & $1\times10^{13}$~n/(cm$^2$ s) & $2.8\times10^{14}$~n/(cm$^2$ s) \\
Fast n Flux    & $5\times10^{12}$~n/cm$^2$ & $1\times10^{12}$~n/(cm$^2$ s) & $7\times10^{12}$~n/(cm$^2$ s)   \\
\hline
\end{tabular}
\end{table}

\section{Assay Results}\label{Sec:Results}
The results of our assay program are listed in Table~\ref{tab:assay}. This Table lists the fractional content of K, U and Th contained within the numerous materials and products assayed during development and assembly of the \DEM. In addition, a number of parts or processes warrant special description due to the techniques developed for their assay. In this section we discuss such developmental work in a series of subsections.

\begin{landscape}
\begin{longtable}{rlcccc}
\caption{Radioactive isotope levels within various materials and their 68\% CL uncertainties. Values for K were not always provided by the analysis.} \\
\label{tab:assay}
\small

\# & Material & Method & K ($10^{-9}$ g/g) & \nuc{232}{Th} ($10^{-12}$ g/g)  & \nuc{238}{U} ($10^{-12}$ g/g)    \\ 
\hline
\endhead
&{\em Metals} &  &  &  &    \\ 
\refstepcounter{rownumber}\label{row:r1} \arabic{rownumber} & Cu Electroformed   Stock Sample & ICPMS &  & $<$0.17 &     \\ 
\refstepcounter{rownumber}\label{row:r5} \arabic{rownumber}  & Cu Electroformed  Stock Sample & ICPMS &  & 0.011$\pm$0.005 & 0.017$\pm$0.003   \\ 
\refstepcounter{rownumber}\label{row:r300} \arabic{rownumber}  & Cu Electroformed  Stock Sample & GDMS & $<$2.2 & $<$50 & $<$70  \\

\refstepcounter{rownumber}\label{row:r200} \arabic{rownumber}  & Cu Electroformed Stock Sample & ICPMS & & 		$<$0.029	& 		$<$0.008	   \\  
\refstepcounter{rownumber}\label{row:r201} \arabic{rownumber}  & Cu Electroformed Stock Sample   & ICPMS &  & 		$<$0.029	& 		$<$0.009	   \\ 
\refstepcounter{rownumber}\label{row:r202} \arabic{rownumber}  & Cu Electroformed Stock Sample  & ICPMS &  & 		$<$0.029	& 		$<$0.008	   \\ 
\refstepcounter{rownumber}\label{row:r203} \arabic{rownumber}  & Cu Electroformed Stock Sample  & ICPMS &  & 		$<$0.030	& 		$<$0.009	   \\ 

\refstepcounter{rownumber}\label{row:r204} \arabic{rownumber}  & Cu Electroformed, machined part, guide clip  & ICPMS &  & 	$0.330\pm0.022$		& 	$0.123\pm0.005$		   \\ 
\refstepcounter{rownumber}\label{row:r205} \arabic{rownumber}  & Cu Electroformed, machined part, guide clip & ICPMS &  & 	$0.112\pm0.009$		& 	$0.078\pm0.002$		   \\ 
\refstepcounter{rownumber}\label{row:r206} \arabic{rownumber}  & Cu Electroformed, machined part, guide clip  & ICPMS & & 	$0.170\pm0.008$		& 	$0.087\pm0.002$		   \\ 

\refstepcounter{rownumber}\label{row:r207} \arabic{rownumber}  & Cu Electroformed, machined part, spring clip  & ICPMS &  & 	$0.215\pm0.009$		& 	$0.130\pm0.010$		   \\ 

\refstepcounter{rownumber}\label{row:r208} \arabic{rownumber}  & Cu Electroformed, machined part, hex bolt      & ICPMS &  & 	$0.118\pm0.011$		& 	$0.035\pm0.004$		   \\ 
\refstepcounter{rownumber}\label{row:r209} \arabic{rownumber}  & Cu Electroformed, machined part, hex bolt     & ICPMS & & 	$0.119\pm0.014$		& 	$0.041\pm0.003$		   \\ 
\refstepcounter{rownumber}\label{row:r210} \arabic{rownumber}  & Cu Electroformed, machined part, hex bolt      & ICPMS &  & 	$0.148\pm0.021$		& 	$0.051\pm0.002$		   \\ 

\refstepcounter{rownumber}\label{row:r218} \arabic{rownumber}  & Cu, C10100 cake stock, (source for Rows~\ref{row:r3},\ref{row:r4})          & ICPMS &  & 	$0.46\pm0.06$ 		& 	$0.21\pm0.06$		   \\ 
\refstepcounter{rownumber}\label{row:r3} \arabic{rownumber} &  Cu, C10100 2.5" plate stock, exterior sample  & ICPMS &  						& 0.27$\pm$0.05 & 0.10$\pm$0.02   \\    
\refstepcounter{rownumber}\label{row:r4} \arabic{rownumber}  &  Cu, C10100 2.5" plate stock,  interior sample & ICPMS &  						& 0.27$\pm$0.05 & 0.12$\pm$0.02   \\ 

\refstepcounter{rownumber}\label{row:r215} \arabic{rownumber}  & Cu, C10100 1" plate stock, saw cut (same stock Row~\ref{row:r216})                                     & ICPMS &  & 	$10.2\pm1.0$		& 	$6.62\pm0.58$		   \\ 
\refstepcounter{rownumber}\label{row:r216} \arabic{rownumber}  & Cu, C10100 1" plate stock, machined surfaces                                          & ICPMS &  & 	$1.88\pm0.45$		& 	$3.11\pm0.39$		   \\  
\refstepcounter{rownumber}\label{row:r217} \arabic{rownumber}  & Cu, C10100 1" x 2" bar stock, machined surfaces         & ICPMS &  & 	$2.12\pm0.39$		& 	$2.25\pm0.15$		   \\  

\refstepcounter{rownumber}\label{row:r211} \arabic{rownumber}  & Cu, C10100 1" plate stock  & ICPMS &  & 	$<$0.029		& 	$0.013\pm0.002$		   \\   
\refstepcounter{rownumber}\label{row:r212} \arabic{rownumber}  & Cu, C10100 2.5" plate stock  & ICPMS &  & 	$<$0.030		& 	$0.017\pm0.003$		   \\  
\refstepcounter{rownumber}\label{row:r214} \arabic{rownumber}  & Cu, C10100 2.5" plate stock   & ICPMS &  & $0.049\pm0.010$ & 	$0.061\pm0.006$		   \\  
\refstepcounter{rownumber}\label{row:r213} \arabic{rownumber}  & Cu, C10100 0.5" plate stock            & ICPMS &  & 	$<$0.030		& 	$0.009\pm0.001$		   \\  

\refstepcounter{rownumber}\label{row:r6} \arabic{rownumber}  & Cu wire, California Fine Wire & ICPMS & $<$25000 & $<$87 & $<$40   \\ 
\refstepcounter{rownumber}\label{row:r7} \arabic{rownumber}  & Pb, smelted from virgin ore, Sullivan Metals & $\gamma$ count & $<$60 & $<$100 & $<$30   \\ 
\refstepcounter{rownumber}\label{row:r8} \arabic{rownumber}  & Pb, UW  & $\gamma$ count & $<$190 & $<$200 & $<$500   \\ 
\refstepcounter{rownumber}\label{row:r9} \arabic{rownumber}  & Pb, UW  & $\gamma$ count & $<$160 & $<$170 & $<$400   \\ 
\refstepcounter{rownumber}\label{row:r10} \arabic{rownumber}  & Pb, smelted from virgin ore, Sullivan Metals  & $\gamma$ count & $<$160 & $<$173 & $<$241   \\ 
\refstepcounter{rownumber}\label{row:r11} \arabic{rownumber}  & Pb, smelted from virgin ore, Sullivan Metals  & GDMS & 4$\pm$2 & $<$10 & $<$10   \\ 
\refstepcounter{rownumber}\label{row:r12} \arabic{rownumber}  & Pb, UW  & GDMS & 23$\pm$11 & $<$8 & $<$10   \\ 
\refstepcounter{rownumber}\label{row:r13} \arabic{rownumber}  & Pb, UW  & GDMS & $<$0.4 & $<$8 & $<$9   \\ 
\refstepcounter{rownumber}\label{row:r14} \arabic{rownumber}  & Pb, archeological ingot, UChicago & GDMS & $<$0.3 & $<$9 & $<$9   \\ 
\refstepcounter{rownumber}\label{row:r15} \arabic{rownumber}  & Pb, archeological sample prepared by Mifer Brick & GDMS & $<$0.2 & $<$8 & $<$10   \\ 
\refstepcounter{rownumber}\label{row:r16} \arabic{rownumber}  & Pb (Average from Brick samples) & ICPMS &  & $1.3\pm1.3$ & $2.9\pm2.0$   \\ 
\refstepcounter{rownumber}\label{row:r19} \arabic{rownumber}  & Sn, sample of unknown origin & $\gamma$ count & 800$\pm$450 & $<$760 & $<$137   \\ 
\refstepcounter{rownumber}\label{row:r20} \arabic{rownumber}  & Sn, sample of unknown origin  & ICPMS & $<$108 & 940$\pm$50 & 1190$\pm$170   \\ 
\refstepcounter{rownumber}\label{row:r21} \arabic{rownumber}  & Sn, sample supplied by Canberra  & ICPMS & $<$108 & 760$\pm$70 & 1150$\pm$350   \\ 
\refstepcounter{rownumber}\label{row:r22} \arabic{rownumber}  & 6-way SS conflat intersection, MDC Vac. Prod., LLC & $\gamma$ count & $<$840 & 3200$\pm$1000 & $<$400   \\ 
\refstepcounter{rownumber}\label{row:r25} \arabic{rownumber}  & TIG-Ce welding rods & $\gamma$ count &$(1.60\pm0.14)\times10^{5}$  &$(1.68\pm0.31)\times10^{7}$ & $<$72000   \\ 
\refstepcounter{rownumber}\label{row:r26} \arabic{rownumber}  & TIG-Zr welding rods & $\gamma$ count & 5500$\pm$4600 &$(1.08\pm0.05)\times10^{5}$  & 19300$\pm$1600   \\ 
\refstepcounter{rownumber}\label{row:r23} \arabic{rownumber}  & Cr, stock used for vapor depositions & ICPMS & $<$7000 & $<$20000 & $<$5000   \\ 
\refstepcounter{rownumber}\label{row:r24} \arabic{rownumber}  & Au, sputtering target & ICPMS & $<$270 &  & 570$\pm$130   \\ 

\refstepcounter{rownumber}\label{row:r604} \arabic{rownumber}  & Au (4N8), sputtered at LBNL & ICPMS & 47000$\pm$1000 & $1980\pm370$ & 2000$\pm$300   \\

\refstepcounter{rownumber}\label{row:r27} \arabic{rownumber}  & Al, sputtered, sample film provided by ORTEC & ICPMS &$(1.42\pm0.51)\times10^{7}$  & 2000$\pm$250 & 5730$\pm$300   \\ 
\refstepcounter{rownumber}\label{row:r28} \arabic{rownumber}  & Al, sputtered, sample film provided by ORTEC & ICPMS & $(1.10\pm0.01)\times10^{5}$ & 2210$\pm$460 & 4390$\pm$340   \\ 
\refstepcounter{rownumber}\label{row:r29} \arabic{rownumber}  & Ge, sputtered, sample film provided by ORTEC& ICPMS & $<$430 & 207$\pm$38 & 843$\pm$62   \\ 
\refstepcounter{rownumber}\label{row:r30} \arabic{rownumber}  & Ge, sputtered, sample film provided by ORTEC & ICPMS & $<$215 & 349$\pm$80 & 1340$\pm$120   \\ 

\refstepcounter{rownumber}\label{row:r603} \arabic{rownumber}  & amorphous Ge, sputtered at LBNL & ICPMS & $4800\pm230$  & 2370$\pm$690 & 1680$\pm$350   \\ 
\refstepcounter{rownumber}\label{row:r603} \arabic{rownumber}  & Cr, sputtered at LBNL & ICPMS & $<$1900  & 5240$\pm$1290 & 5030$\pm$700   \\ 

\refstepcounter{rownumber}\label{row:r605} \arabic{rownumber}  & Ti film, sputtered at LBNL & ICPMS &   & $<$400 & $<$100   \\

\\
& {\em Plastics} &  &  &  &    \\ 

\refstepcounter{rownumber}\label{row:r35} \arabic{rownumber}  & Teflon\Rmark\ TE-6742 & NAA & 0.15$\pm$0.02 & 0.025$\pm$0.002 & $<$0.4   \\ 
\refstepcounter{rownumber}\label{row:r34} \arabic{rownumber}  & Peek\Rmark, Victrex\Rmark\ & NAA & 180$\pm$110 & $<$400 & $<$5100   \\ 
\refstepcounter{rownumber}\label{row:r36} \arabic{rownumber}  & Vespel\Rmark, Dupont\Tmark\ rod, ThyssenKrupp Materials & NAA & 350$\pm$300 & $<$2.9 & $<$84.4   \\ 
\refstepcounter{rownumber}\label{row:r37} \arabic{rownumber}  & Vespel\Rmark, Dupont\Tmark\ rod, Professional Plastics Inc. & NAA & 191$\pm$31 & $<$49 & $<$45   \\ 
\refstepcounter{rownumber}\label{row:r38} \arabic{rownumber}  & Vespel\Rmark, Dupont\Tmark\ plate, Professional Plastics Inc. & NAA & $<$200 & $<$42 & $<$930   \\ 
\refstepcounter{rownumber}\label{row:r31} \arabic{rownumber}  & Parylene N dimer, Para Tech Coating Inc. & NAA &  & $<$50 & $<$30   \\ 
\refstepcounter{rownumber}\label{row:r47} \arabic{rownumber}  & Parylene, Speciality Coating Systems\Tmark\ Inc.  &  NAA  &  5800$\pm$1300  &  $<$850  &  $<$1700   \\ 
\refstepcounter{rownumber}\label{row:r40} \arabic{rownumber}  & Parylene C, dimer, Speciality Coating Systems\Tmark\ Inc. & ICPMS & 2110$\pm$15 & 390$\pm$30 & 6230$\pm$110   \\ 
\refstepcounter{rownumber}\label{row:r41} \arabic{rownumber}  & Parylene C, dimer pre cleaned, Spec. Coat. Sys. Inc. & ICPMS & $<$108 & 37$\pm$3 & 4230$\pm$60   \\ 
\refstepcounter{rownumber}\label{row:r42} \arabic{rownumber}  & Parylene C, infusion rod bump, Spec. Coat. Sys.  Inc. & ICPMS & $<$320 & 250$\pm$60 & 46$\pm$20   \\ 
\refstepcounter{rownumber}\label{row:r43} \arabic{rownumber}  & Parylene C, inlet, Speciality Coating Systems\Tmark\ Inc. & ICPMS & $<$430 & 140$\pm$30 & 83$\pm$24   \\ 
\refstepcounter{rownumber}\label{row:r44} \arabic{rownumber}  & Parylene C, table, Speciality Coating Systems\Tmark\ Inc. & ICPMS & 923$\pm$86 & 530$\pm$30 & 250$\pm$60   \\ 
\refstepcounter{rownumber}\label{row:r45} \arabic{rownumber}  & 5\% boron HDPE, Plasti-Shield\Rmark, King Plastic  & $\gamma$ count & $<$8000 & $<$8000 & $<$3000   \\ 
\refstepcounter{rownumber}\label{row:r46}  \arabic{rownumber} & Densetec HDPE, Polymer Industries   & $\gamma$ count & $1400\pm200$ & $3000\pm200$ & $1600\pm100$  \\ 
\refstepcounter{rownumber}\label{row:r49} \arabic{rownumber}  & PTFE, 0.5" tubing, Cole-Palmer & ICPMS &  & 1.5$\pm$1.8 & $<$3.1   \\ 
\refstepcounter{rownumber}\label{row:r50} \arabic{rownumber}  & Kalrez\Rmark, Dupont\Tmark, O-ring material & NAA &  & 4700$\pm$200 & $<$4100   \\ 
\refstepcounter{rownumber}\label{row:r51} \arabic{rownumber}  & Kalrez\Rmark, Dupont\Tmark, O-ring material & ICPMS &  & 1785$\pm$890 & 754$\pm$57   \\ 
\refstepcounter{rownumber}\label{row:r54} \arabic{rownumber}  & FEP shrink tubing ACCU-GLASS & NAA & $<$19 & $<$34 & $<$100   \\ 
\refstepcounter{rownumber}\label{row:r55} \arabic{rownumber}  & FEP shrink tubing NEWARK & NAA & 20.6$\pm$2.3 & $<$12 & 8.0$\pm$3.4   \\ 
\refstepcounter{rownumber}\label{row:r301} \arabic{rownumber}  & FEP tubing  & NAA &    & $<$45 & $<$150   \\ 
\refstepcounter{rownumber}\label{row:r302} \arabic{rownumber}  & PTFE 0.002-in sheet & NAA &   & $<$5.1 & $<$7.6   \\ 
\\
& {\em Cables and parts} &  &  &  &    \\ 

\refstepcounter{rownumber}\label{row:r52} \arabic{rownumber}  & FEP, Dupont\Tmark\ NP20, outer jacket Axon' & NAA & $<$1100 & $<$28 & $<$170   \\ 
\refstepcounter{rownumber}\label{row:r53} \arabic{rownumber}  & FEP, Dupont\Tmark\ NP20, inner dielectric Axon'  & NAA & $<$3700 & $<$27 & $<$140   \\ 
\refstepcounter{rownumber}\label{row:r56} \arabic{rownumber}  & Axon' Picocoax\Rmark, Model PCX46P10EP, 0.12 g/m & $\gamma$ count & 22400$\pm$6400 & 14000$\pm$2000 & $<$28000   \\ 
\refstepcounter{rownumber}\label{row:r57} \arabic{rownumber}  & Axon' Picocoax\Rmark, same as row~\ref{row:r56} & ICPMS &  & $<2\times10^{6}$ & $<8\times10^{5}$   \\ 
\refstepcounter{rownumber}\label{row:r58} \arabic{rownumber}  & Cu wire, bare AWG34, MWS Wire Industries & ICPMS & $<$320 & 33.2$\pm$8.9 & 26.6$\pm$10.1   \\ 
\refstepcounter{rownumber}\label{row:r59} \arabic{rownumber}  & Cu wire, bare AWG40, MWS Wire Industries & ICPMS & $<$270 & 38.9$\pm$8.4 & 14.1$\pm$12.5   \\ 
\refstepcounter{rownumber}\label{row:r60} \arabic{rownumber}  & Mini Coax cable, Cooner Wire Inc. & ICPMS & & $<2\times10^{6}$ & $<8\times10^{5}$     \\ 
\refstepcounter{rownumber}\label{row:r62} \arabic{rownumber}  & Mini Coax cable, Cooner Wire Inc.  & $\gamma$ count & 11600$\pm$5000 & $<$4000 & $<$21000   \\ 
\refstepcounter{rownumber}\label{row:r65} \arabic{rownumber}  & Handmade Parylene coated Cu wire & ICPMS & & 815$\pm$81 & 330$\pm$33   \\ 
\refstepcounter{rownumber}\label{row:63r} \arabic{rownumber}  & Axon' Picocoax\Rmark, same as row~\ref{row:r56} & $\gamma$ count & 21000$\pm$2000 & $<$8000 & 3000$\pm$1000   \\ 
\refstepcounter{rownumber}\label{row:r64} \arabic{rownumber}  & Axon' Picocoax\Rmark, same as row~\ref{row:r56} & ICPMS &  & 566$\pm$57 & 3100$\pm$310   \\ 
\refstepcounter{rownumber}\label{row:r66} \arabic{rownumber}  & Axon'  Picocoax\Rmark, 3 g/m, custom HV cable & ICPMS &  & 470$\pm$110 & 5900$\pm$300   \\ 
\refstepcounter{rownumber}\label{row:r67} \arabic{rownumber}  & Axon'Picocoax\Rmark,  0.4 g/m custom LV cable & ICPMS &  & 220$\pm$71 & 940$\pm$37   \\ 
\refstepcounter{rownumber}\label{row:r70} \arabic{rownumber}  & Axon'  Picocoax\Rmark, 3 g/m, custom HV cable & ICPMS &  & 0.54$\pm$0.05 & 11.7$\pm$1.2   \\ 	
\refstepcounter{rownumber}\label{row:r72} \arabic{rownumber}  & Axon'Picocoax\Rmark,  0.4 g/m custom LV cable & $\gamma$ count & $<$400 & $<$700 & $<$200   \\ 
\refstepcounter{rownumber}\label{row:r73} \arabic{rownumber} & Axon' Signal Cable (final) & $\gamma$ count & $<$2000 & $<$1000 & $<$800   \\ 
\\
 & {\em Connectors, front-end electronics, and small parts} &  &  &   &   \\ 

\refstepcounter{rownumber}\label{row:r75} \arabic{rownumber}  & Silver epoxy part 1, TRA-DUCT 2902, Tra-Con, Inc.& ICPMS & $<$200 & 330$\pm$4 & 64$\pm$4   \\ 
\refstepcounter{rownumber}\label{row:r76} \arabic{rownumber}  & Silver epoxy part 2, TRA-DUCT  2902, Tra-Con, Inc.& ICPMS & $<$200 & 578$\pm$24 & 349$\pm$12   \\ 
\refstepcounter{rownumber}\label{row:r77} \arabic{rownumber}  & Silver epoxy part 1, TRA-DUCT 2902, Tra-Con, Inc.& ICPMS & 107$\pm$4 & 26.4$\pm$4.4 & 20$\pm$3   \\ 
\refstepcounter{rownumber}\label{row:r78} \arabic{rownumber}  & Silver epoxy part 2, TRA-DUCT  2902, Tra-Con, Inc.& ICPMS & 518$\pm$10 & 56.7$\pm$7 & 67$\pm$8   \\ 
\refstepcounter{rownumber}\label{row:r79} \arabic{rownumber}  & Silver epoxy, TRA-DUCT 2902, Tra-Con, Inc. & $\gamma$ count & $<$30000 & $<$70000 & $<$10000   \\ 
\refstepcounter{rownumber}\label{row:r80} \arabic{rownumber}  & Silver epoxy packs, TRA-DUCT 2902, Tra-Con, Inc.& $\gamma$ count & 84000$\pm$8000 & $<$8000 & $<$4000   \\ 
\refstepcounter{rownumber}\label{row:r81} \arabic{rownumber}  & Silver epoxy, TRA-DUCT 2902, Tra-Con, Inc.& $\gamma$ count &  $<$3000  & $<$5000 & $20000\pm3000$   \\ 
\refstepcounter{rownumber}\label{row:r82} \arabic{rownumber}  & Silver epoxy hard., TRA-DUCT 2902, Tra-Con, Inc. & ICPMS & 556$\pm$11 & 56.7$\pm$7 & 67$\pm$8   \\ 
\refstepcounter{rownumber}\label{row:r87} \arabic{rownumber}  & Silver epoxy, TRA-DUCT 2902, Tra-Con, Inc. & ICPMS & $<$70 & 79$\pm$4 & 11$\pm$1   \\ 
\refstepcounter{rownumber}\label{row:r85} \arabic{rownumber}  & Silver Epoxy, Emerson-Cumings & $\gamma$ count & $<$2000 & $<$3000 & $<$1000   \\ 
\refstepcounter{rownumber}\label{row:r400} \arabic{rownumber}  & SnAg Solder, NCD			& GDMS	& $<$73	&$<$70	&$<$200	\\
\refstepcounter{rownumber}\label{row:r500} \arabic{rownumber}  & Indium Corp. Solder			& GDMS	& $<$120	&$<$100	&$<$300	\\
\refstepcounter{rownumber}\label{row:r83} \arabic{rownumber}  & Abietic acid, tech grade, Sigma Aldrich & ICPMS & $<$108 & $<$90 & 93$\pm$16   \\ 
\refstepcounter{rownumber}\label{row:r84} \arabic{rownumber}  & Abietic acid, tech grade, Sigma Aldrich & NAA & $<$28 & $<$28 & $<$60   \\ 
\refstepcounter{rownumber}\label{row:r86} \arabic{rownumber}  & Soap solution, Micro-90\Rmark\  & $\gamma$ count & 42000$\pm$2000 & $<$1500 & $<$600   \\ 
\refstepcounter{rownumber}\label{row:r74} \arabic{rownumber}  & Soap solution, Liquinox\Rmark\ Alconox Inc. & $\gamma$ count & $1\times10^{7}$ & $<$10000 & $<$2000   \\ 
\refstepcounter{rownumber}\label{row:r88} \arabic{rownumber}  & Fused silica, Corning 7980 ArF & $\gamma$ count & $<$90000 & $<$70000 & $<$40000   \\ 
\refstepcounter{rownumber}\label{row:r126} \arabic{rownumber}  & Fused silica, Corning 7980 KrF & ICPMS &  & $187\pm26$& $309\pm69$ \\
\refstepcounter{rownumber}\label{row:r127} \arabic{rownumber}  & Fused silica, Corning 7980 KrF & ICPMS &  & $74\pm23$& $99\pm12$ \\
\refstepcounter{rownumber}\label{row:r128} \arabic{rownumber}  & Fused silica, Corning 7980  std. polish & ICPMS &  &$97\pm41$ & $54\pm20$ \\
\refstepcounter{rownumber}\label{row:r129} \arabic{rownumber}  & Fused silica, Corning 8650 ArF & ICPMS &  & $118\pm25$ & $525\pm134$ \\

\refstepcounter{rownumber}\label{row:r602a} \arabic{rownumber}  & Fused silica, Corning 7980 prototype wafer, uncleaned & ICPMS &  & $11\pm5$ & $160\pm20$ \\
\refstepcounter{rownumber}\label{row:r602} \arabic{rownumber}  & Fused silica, Corning 7980 prototype wafer, cleaned & ICPMS &  & $<$10 & $100\pm10$ \\

\refstepcounter{rownumber}\label{row:r89} \arabic{rownumber}  & CFW Al-Si bonding wire & ICPMS & 3280$\pm$27 & 91000$\pm$2000 & 8700$\pm$400   \\ 
\refstepcounter{rownumber}\label{row:r90} \arabic{rownumber}  & Pins without Beryllium/Copper Contacts & ICPMS & 29000$\pm$530 & 1500$\pm$400 & 310$\pm$70   \\ 
\refstepcounter{rownumber}\label{row:r91} \arabic{rownumber}  & Pins with Beryllium/Copper Contacts & ICPMS & 30000$\pm$510 & 9900$\pm$300 & 64000$\pm$100   \\ 
\refstepcounter{rownumber}\label{row:r92} \arabic{rownumber}  & Vespel\Rmark, in-house machined housing & NAA & $<$935 & 310$\pm$140 & $<$190   \\ 
\refstepcounter{rownumber}\label{row:r93} \arabic{rownumber}  & In-house machined female connector, full body assay & ICPMS &  & 96.2$\pm$1.7 & 43.4$\pm$0.7   \\ 
\refstepcounter{rownumber}\label{row:r94} \arabic{rownumber}  & In-house machined male connector, full body assay & ICPMS &  & 7.1$\pm$0.5 & 10.5$\pm$0.9   \\ 
\refstepcounter{rownumber}\label{row:r95} \arabic{rownumber}  & Sapphire C-Plane, 0.35 mm thick, Marketech Intern. Inc. & NAA & $<$376 & $<$21 & $<$300   \\ 
\refstepcounter{rownumber}\label{row:r96} \arabic{rownumber}  & JFET dies, MX-17A, MOKTEK & ICPMS &  & $<$1900 & $<$140   \\ 

\refstepcounter{rownumber}\label{row:r601} \arabic{rownumber}  & LMFE1, LMFE prototype without JFET & ICPMS &  $386\pm12$ & 130$\pm$10 & 430$\pm$43   \\ 

\refstepcounter{rownumber}\label{row:r97} \arabic{rownumber}  & LMFE2a, Full board, internal ID 1B0109 & ICPMS & $<$215 & 2120$\pm$60 & 850$\pm$30   \\ 
\refstepcounter{rownumber}\label{row:r98} \arabic{rownumber}  & LMFE2b, Full board, internal ID 1B0110 & ICPMS & $<$215 & 1610$\pm$30 & 850$\pm$30   \\ 
\\
 & {\em Miscellaneous} &  &  &   &   \\
 
\refstepcounter{rownumber}\label{row:r99} \arabic{rownumber}  & Precision Urethane Drive Roller (70a durometer) & $\gamma$ count & $<$500000 & $<$175000 & $<$65000   \\ 
\refstepcounter{rownumber}\label{row:r100} \arabic{rownumber}  & Wipes, KIMTECH PURE\Rmark\ W4, Kimberly-Clark Prof.\Rmark\ & $\gamma$ count & 44800$\pm$12800 & (1.00$\pm0.15)\times10^{6}$ & $<$56000   \\ 
\refstepcounter{rownumber}\label{row:r101} \arabic{rownumber}  & Charcoal, 102022, finer size grain, Bl\"{u}cher   & ICPMS & 20400$\pm$170 & 369$\pm$30 & 1870$\pm$70   \\ 
\refstepcounter{rownumber}\label{row:r102} \arabic{rownumber}  & Charcoal, 101135 Saratoga, 0.47 mm, Bl\"{u}cher  & ICPMS & 11300$\pm$46 & 181$\pm$18 & 385$\pm$16   \\ 
\refstepcounter{rownumber}\label{row:r107} \arabic{rownumber}  & Charcoal, 101135 Saratoga, 0.47 mm, Bl\"{u}cher  & $\gamma$ count & 10000$\pm$3400 & 5180$\pm$246 & 5870$\pm$960     \\ 
\refstepcounter{rownumber}\label{row:r103} \arabic{rownumber}  & Charcoal, UHP granules, Carbo-Act Int.& ICPMS & 2880$\pm$12 & 458$\pm$48 & 647$\pm$48   \\ 
\refstepcounter{rownumber}\label{row:r104} \arabic{rownumber}  & Charcoal, sample from MPI, Heidelberg  & ICPMS & 117$\pm$2 & 135$\pm$20 & 606$\pm$9   \\ 		
\refstepcounter{rownumber}\label{row:r116} \arabic{rownumber}  & Charcoal, K48, Silcarbon  & ICPMS & $(1.17\pm0.01)\times10^{7}$ & 4070$\pm$150 & 3990$\pm$80     \\ 
\refstepcounter{rownumber}\label{row:r105} \arabic{rownumber}  & Charcoal, Calgon Carbon  & $\gamma$ count & $(6.40\pm0.14)\times10^{5}$ & $(2.610\pm0.079)\times10^{5}$ & 83000$\pm$2700   \\ 
\refstepcounter{rownumber}\label{row:r106} \arabic{rownumber}  & Charcoal, source from Canberra & $\gamma$ count & $(1.50\pm0.08)\times10^{5}$ & $(1.545\pm0.055)\times10^{5}$ & 18800$\pm$1700     \\ 
\refstepcounter{rownumber}\label{row:r108} \arabic{rownumber}  & Hysol\Rmark\  0151\Tmark\ resin, McMaster-Carr & ICPMS & $<$86 & 130$\pm$10 & 17$\pm$5     \\ 
\refstepcounter{rownumber}\label{row:r109} \arabic{rownumber}  & Hysol\Rmark\  0151\Tmark\ hardener, McMaster-Carr  & ICPMS & $<$215 & 140$\pm$20 & 39$\pm$7     \\ 
\refstepcounter{rownumber}\label{row:r110} \arabic{rownumber}  & Hysol\Rmark\  1C\Tmark\ resin, McMaster-Carr  & ICPMS & 6200$\pm$200 & 3190$\pm$290 & 86300$\pm$1100     \\ 
\refstepcounter{rownumber}\label{row:r111} \arabic{rownumber}  & Hysol\Rmark\  1C\Tmark\ hardener, McMaster-Carr  & ICPMS & 26670$\pm$260 & 79200$\pm$800 & 1560$\pm$1000     \\ 
\refstepcounter{rownumber}\label{row:r112} \arabic{rownumber}  & Torr Seal\Rmark\  Base, McMaster-Carr  A & ICPMS & $<$7530 & 2350$\pm$480 & 10300$\pm$2000     \\ 
\refstepcounter{rownumber}\label{row:r113} \arabic{rownumber}  & Torr Seal\Rmark\  Hardener B, McMaster-Carr  & ICPMS & 35400$\pm$250 & 45200$\pm$700 & 148000$\pm$2000     \\ 
\refstepcounter{rownumber}\label{row:r114} \arabic{rownumber}  & Silicone Rubber, P-4, Silicones Inc. & $\gamma$ count & $<$2000 & $<$3000 & $<$6000     \\ 
\refstepcounter{rownumber}\label{row:r115} \arabic{rownumber}  & 2-propanol, A461-500, Fischer Scientific & ICPMS & $<$10 & $<$0.1 & $<$0.1     \\ 
\refstepcounter{rownumber}\label{row:r117} \arabic{rownumber}  & Mix colored beads, 100780, Accu-Glass Prod. Inc. & ICPMS & $(2.63\pm0.01)\times10^{7}$ & 39500$\pm$2000 & 453000$\pm$6300     \\ 
\refstepcounter{rownumber}\label{row:r118} \arabic{rownumber}  & White beads, 100780, Accu-Glass Products Inc.  & ICPMS & $(2.82\pm0.01)\times10^{7}$ & 40200$\pm$1900 & 713000$\pm$12000     \\ 
\refstepcounter{rownumber}\label{row:r119} \arabic{rownumber}  & Green beads, 100780, Accu-Glass Products Inc.  & ICPMS & $(3.05\pm0.01)\times10^{7}$ & $(2.23\pm0.08)\times10^{5}$ & 386000$\pm$8000      \\ 
\refstepcounter{rownumber}\label{row:r120} \arabic{rownumber}  & Black bead leachate & ICPMS & 1890$\pm$10 & 57.2$\pm$2.2 & 43$\pm$2     \\ 
\refstepcounter{rownumber}\label{row:r121} \arabic{rownumber}  & Blue bead leachate & ICPMS & 7850$\pm$41 & 150$\pm$3 & 259$\pm$4     \\ 
\refstepcounter{rownumber}\label{row:r122} \arabic{rownumber}  & Brown bead leachate & ICPMS & 2030$\pm$9 & 64.7$\pm$3 & 46.5$\pm$1.7   \\ 
\refstepcounter{rownumber}\label{row:r123} \arabic{rownumber}  & Green bead leachate & ICPMS & 3820$\pm$20 & 261$\pm$4 & 195$\pm$5   \\ 
\refstepcounter{rownumber}\label{row:r124} \arabic{rownumber}  & Grey bead leachate & ICPMS & 2120$\pm$16 & 281$\pm$6 & 199$\pm$3   \\ 
\refstepcounter{rownumber}\label{row:r125} \arabic{rownumber}  & White bead leachate & ICPMS & 2460$\pm$12 & 67$\pm$2 & 59$\pm$1   \\ 
\hline

\end{longtable}

\end{landscape}

\subsection{Underground Electroformed Copper}
The \DEM\ is constructed with a large quantity of copper used in the cryostat and inner shield. As a result, it has a significant impact on the background model and requires special consideration. Electroforming Cu is effective at removing impurities~\cite{Hoppe2014,Hoppe2008,Hoppe2009}. Electroforming and machining the Cu underground prevents the ingrowth of cosmogenically produced isotopes. While we were able to locate one lot of commercial copper that was adequately pure to obtain non-detect values using the ICPMS assay, this was premachined, heavily-etched, bulk material.  Our search for such clean commercial copper underscores the wide range of U and Th contamination found in such materials and how critical any subsequent handling is for maintaining its purity.

Early on, the \MJ\ collaboration determined copper to be the most desirable material for detector string parts and cryostat construction due to its favorable thermal, electrical, and mechanical properties~\cite{Overman2012,Overman2015,Overman2015A}.  Although it can be obtained commercially with relatively high purity, it is contaminated by a variety of means.  During the production and subsequent handling processes contamination with U and Th can occur.  Copper also suffers from activation due to reactions with cosmic-ray produced secondary neutrons, e.g., \nuc{63}{Cu}(n,$\alpha$)\nuc{60}{Co}.  \nuc{60}{Co} is long-lived (half-life = 5.3 yr) and attempting to limit this activation presents its own challenges.  These considerations motivate underground electroforming of the \MJ\ \DEM\ copper, to not only produce Cu below the target purity level of 0.3 $\mu$Bq/kg for \nuc{238}{U} and \nuc{232}{Th}, but also to minimize the cosmic ray exposure.

Copper for the \MJ\ \DEM\ was electroformed in two underground locations. Seven electrochemical baths operated at the Shallow Underground Laboratory (SUL) at PNNL with an overburden of $\sim$40 feet. Ten baths operated in a dedicated clean room facility constructed at the 4850 ft level at SURF. Construction of the electroforming baths used parts selected for their long-term compatibility with sulfuric acid and copper sulfate solutions. Since the bath components were one potential source of radionuclide contaminants, background mitigation required careful design and construction. Excluding the copper anode material, the only materials permitted to contact the bath solution were acid-leached polymers. The bath materials include polypropylene, high-density polyethylene, and polytetrafluoroethylene. All of these materials underwent a rigorous cleaning and high-purity acid-leaching process to remove contaminants~\cite{Overman2013}. These techniques required the use of greater than 99.995\% pure oxygen free high conductivity (OFHC) copper anodes, which are commercially available. We purchased ours from Titan International~\cite{TitanCu} The bath electrolyte was produced using $>$18 M$\Omega$-cm deionized water and Optima\Tmark\ grade sulfuric acid.  The copper sulfate was produced in-house using pure acid and copper in order to achieve the desired purity because no vendor for such high purity material could be found. The gas volume over the bath solution was purged with liquid nitrogen boil-off gas to mitigate radon intrusion.

\MJ\ used potential limited, reverse-pulse electroplating techniques.  To maximize the control over the electrical parameters such as potential and waveform, the electroforming power supplies used in our process were designed under the Department of Energy's Small Business Innovation Research program with Dynatronix, Inc.. This in turn influences purity and the material's physical properties.  The copper purity required production rooms operated in a clean room environment class 1000 or better.

The material can be produced centimeters thick and used as-is. Its mechanical properties~\cite{Overman2012,Overman2015,Overman2015A} are adequate and similar to commercially available copper without requiring additional metallurgical treatment. It thus avoids the introduction of contamination typical of further industrial processing.  

In the absence of a non-destructive direct assay of electroformed copper with adequate sensitivity for measuring Th and U to the required level, the assay of the bath electrolyte was used to infer the material purity.  The concentration of contaminant species such as U and Th in the electrolyte can be used to predict their concentration in the deposited copper. 
 Data obtained previously~\cite{Hoppe2008,Hoppe2009} indicated that the concentration of U and Th as contaminants deposited in electroformed copper are about 1000 pg/g when the concentration in the electrolyte solution was 1000 ng/ml.  Data~\cite{Hoppe2014}, obtained later at a much lower concentration very near the detection limit indicated a comparable albeit decreased comtamination-rejection rate. From these data we determined maximum allowable electrolyte U and Th concentrations for \DEM| copper of 0.024 ng/ml and 0.075 ng/ml, respectively. Electrolyte contamination below these concentration levels indicates that electroformed copper is being produced with impurity levels below 0.3 $\mu$Bq/kg and thereby surpassing the \DEM\ radiopurity goals for U and Th in copper.  The direct assay results of electroformed Cu samples support this conclusion.

To ensure a stronger rejection rate for the \DEM-produced electroformed copper than that measured in Ref.~\cite{Hoppe2014}, much greater care in the control and monitoring of the electroforming processes was implemented.  The electroforming baths used for the \MJ\ \DEM, given their large bulk volume, were also expected to achieve a better rejection rate than the small baths used previously. Contaminants were measured in the electrolyte before electroforming began and then at intervals during deposition.  For example, early concentrations in the baths at PNNL were an average of 0.006 ng/ml U and 0.022 ng/ml Th.  After nearly two months of electroforming, the levels had increased to 0.010 ng/ml U and 0.040 ng/ml Th due to the contributions of the dissolving copper anode material. These increased levels were still below their maximum allowable levels.  The change is proportionately the same for both U and Th, which supports the argument that the source of contamination is from the dissolution of anode copper and not some other outside source.  Since the change is also proportional to the anode stock consumption, it indicates a strong rejection of U and Th. The concentrations of U and Th increasing in the baths were also within expected levels. This verifies that sufficiently high-purity acids were used for the periodic replenishment of the bath electrolyte as required due to bulk chemistry and trace constituent limitations. 

The latest assay results for electroformed copper, presented in Table~\ref{tab:assay}, validate the approach of monitoring the electrolyte contaminant levels to determine the Cu contamination. The new results from this work and Ref.~\cite{LaFerriere2015} are presented in Table~\ref{tab:assay} in Rows~\ref{row:r1} and \ref{row:r5} where Row~\ref{row:r300} is an early assay of that material. Rows~\ref{row:r200} through \ref{row:r203} represent bulk samples from electroformed Cu used within the apparatus. These samples underwent a heavy etch with little or no additional handling. The samples listed in Rows~\ref{row:r204} through \ref{row:r210} represent parts produced for the \DEM\ but sacrificed for assay. These were typical parts from the production line prepared as all parts were. These parts were machined and then etched. The additional handling resulted in a higher level of contamination than in the raw stock material. The numbers are listed in the table as bulk contaminations, but are more consistent with surface contamination at the level of 2.8 $\mu$Bq/m$^2$ and 1.8 $\mu$Bq/m$^2$ for \nuc{238}{U} and \nuc{232}{Th}, respectively. These values were estimated from analyzing samples from various etch depths. Heavy etching results in the ultimate purity level, but part tolerance must be considered. This situation is an area of future research.

\subsection{Commercially Available Copper}
Commercially available copper is used as shielding material in the \DEM\ outside of the electroformed Cu shield. The majority of the commercial Cu is in the form of plates machined to a thickness of 2" as the outer Cu shield between the electroformed inner Cu shield and the Pb shield. Additionally, commercial Cu parts are in other locations where the detector module interfaces with the shield, such as a support stand to support the detector module within the Pb shield. It is also used as filler shielding throughout the penetrations of the Pb shield. 

Several assays were performed during the selection and acquisition of the commercial Cu. With one exception, all commercial Cu assays represent Cu stock purchased through Southern Copper \& Supply Company (USA), which sourced the Cu plate material from KME (Europe) where it is rolled to the desired plate thickness. The original Cu cake material is supplied to KME by either Aurubis or Mitsubishi Materials. 

Samples of the starting cake material (prior to rolling) originating from Aurubis and the final 2.5" rolled plate were assayed prior to the purchase of the final material. Further, separate samples of the rolled plate were collected on the exterior rolled surface (a potential source of contamination) and the interior of the plate. The results of the pre-rolled cake sample (see Row~\ref{row:r218} of Table~\ref{tab:assay}) and the plate samples (see Rows~\ref{row:r3} and \ref{row:r4} of Table~\ref{tab:assay}) show a difference of a factor of 2, which we consider to be of no significance and likely due to sampling issues.  In any case, it is an acceptable radiopurity for the outer Cu shield.  

The detector module is supported within a removable portion of the Pb shield with a Cu support stand made from 1"x1" and 1"x2" Cu bars. The stock material was again plate stock cut to the desired width by the vendor. Once underground in our machine shop, the rough cut bar stock underwent a surface machining on all surfaces and was then cut to length. An initial assay of the 1" plate stock originating from Mitsubishi Materials (See Row~\ref{row:r215}) prior to our own surface machining showed undesirable purity levels, though it was noted that surface preparations for the ICPMS assay were not sufficient to remove the evidence of saw cuts on the surface of the sample. A repeat assay of the same stock (Row~\ref{row:r216} of Table~\ref{tab:assay}) with the production machined surface did show improvement in the radio purity levels. A machined-surface sample of 1"x2" bar stock from Copper and Brass Sales (Row~\ref{row:r217} of Table~\ref{tab:assay}) showed similar radio purity levels.

In an attempt to identify cleaner material for the Cu support structure and noting the success of the acceptable purity found for the outer Cu shield 2.5" plate, a second round of plate stock was evaluated. Since the final 1"x2" portion of the Cu support stand has a more stringent radiopurity requirement, we evaluated a sample of 2.5" plate cut to a 1.25" width, finding much improved radiopurity (Row~\ref{row:r212} of Table~\ref{tab:assay}) without any pre-assay surface machining. In addition, a second 1" plate sample originating from Mitsubishi Materials without any pre-assay surface machining (Row~\ref{row:r211} of Table~\ref{tab:assay}) also shows improved radio purity. The two latter samples represent the Cu stock used for the Cu support stand. 

Two remaining Cu sources are evaluated for use where the detector module interfaces with the shield. A sample of 2.5" plate stock cut to 2.5" width originating from Arubis (Table~\ref{tab:assay}, Row~\ref{row:r214}) is used for 2"x4" Cu bricks around a cylindrical crossarm through the Pb shield. A sample of 0.5" plate stock originating from Mitsubishi (Row~\ref{row:r213}) is used as a shielding seal around the cylindrical crossarm.

\subsection{Lead}
The most abundant material, by mass, used in the \DEM\ is lead, which is the dominant component of the shield. Because it is so massive, it is not possible to directly count all the materials used for the shield. Therefore, we relied on sample testing and indirect assays to assess the lead's purity. 

Two sources of Pb bricks (nominally 2"$\times$~4"$\times~$8" ) were used for the \DEM\ shield. One was a new production run from virgin Doe Run Mine Pb formed into Pb bricks by Sullivan Metals, Inc. (Holly Springs, MS) and the other was from a discontinued, low-background counting facility at the University of Washington. Because different sources were used, the history of the UW Pb was not entirely known, and  and because direct assay of all individual bricks was not feasible, we developed a process to verify the purity of the Pb bricks. Initially, direct assay by $\gamma$-ray counting and GDMS at the NRCC (Section~\ref{Sec:GSMS}) of a small number of Pb bricks from both sources provided confidence of the Pb radiopurity. (See Rows~\ref{row:r7} to~\ref{row:r15} in Table~\ref{tab:assay}.) 

To better understand the uniformity of the purity of the Pb, we took advantage of the fact that all the bricks were cleaned prior to delivery to our underground laboratory at a collaboration facility established at Black Hills State University. We directly sampled a selection of bricks during that cleaning process and assayed those samples using ICPMS at PNNL (Section~\ref{Sec:ICPMS}). The result is given in Row~\ref{row:r16} in Table~\ref{tab:assay}.

The bricks were cleaned in a cleanroom by first soaking them in groups of 5 in a pure ACS-grade acetic acid bath for 3-13 minutes depending on the brick condition. The etch bath was replaced when its effectiveness in cleaning the bricks was noticed to decline as evidenced by the average cleaning time. During this soak the bricks were scrubbed with soft plastic brushes for 1-2 minutes. Following the soak, the bricks were rinsed in successive $>$18-M$\Omega$-cm deionized water baths for 20-30 seconds each. Water rinse baths were replaced when the rinse water appeared dirty. Next they were transferred to a bath of nitric acid and peroxide (1-3\% Optima\Tmark\ grade acid used once previously for leaching plastics, and 3\% non-stabilized ACS grade peroxide) for 1-2 minutes followed by 2 successive 20-30 second deionized water rinses. The nitric-peroxide bath was replaced whenever it was observed that bubble formation on the bricks had diminished or that bricks were not emerging from the bath with the appropriate silvery finish. The bricks were then patted dry with cleanroom-grade wipes to remove excess water, followed by a rinse with isopropyl alcohol for 1-2 minutes. The bricks were then air dried before being triple-bagged in 6-mil polyethylene. Finally, the bricks were palletized for transfer to the \DEM\ clean room underground. The acetic acid etch typically removes about 50 $\mu$m of material but can remove up to 250 $\mu$m. The nitric solution etch removes about 13 $\mu$m. In total over 6800 bricks were cleaned at a rate of 50/d. The \DEM\ shield contains approximately 4500 bricks.

Bulk samples from 40 bricks, 20 from each source, were collected after the nitric-peroxide etch. A gouging tool was used to produce these samples with a mass of approximately 5 g each. These bricks were set aside for storage and not used in the shield. The average of the ICPMS measurements obtained at PNNL indicated contamination levels of 1.3 ppt and 2.9 ppt for Th and U respectively. These results meet our requirements for the Pb and are listed in Row~\ref{row:r16} in Table~\ref{tab:assay}.

\subsection{Parylene}
Parylene generally refers to a family of chemical vapor-deposited polymers used in industry as moisture and dielectric barriers. The \mj\ \dem\ makes use of parylene for multiple purposes. All hardware used to construct the inner regions of the cryostats is made from ultra-pure electroformed Cu. Because of the radiopurity requirements of the experiment, the use of oil-based lubricants is not possible. So parylene is used as a dry-film lubricant for copper nuts and bolts to prevent galling. Parylene, in the form of thin sheets, is also used as an insulator for electrical isolation of components of the copper shield and copper cryostat. Assays of parylene products from different manufacturers were performed.

\begin{figure}[htdp]
	\centering
	\includegraphics[height=6cm, width=4cm]{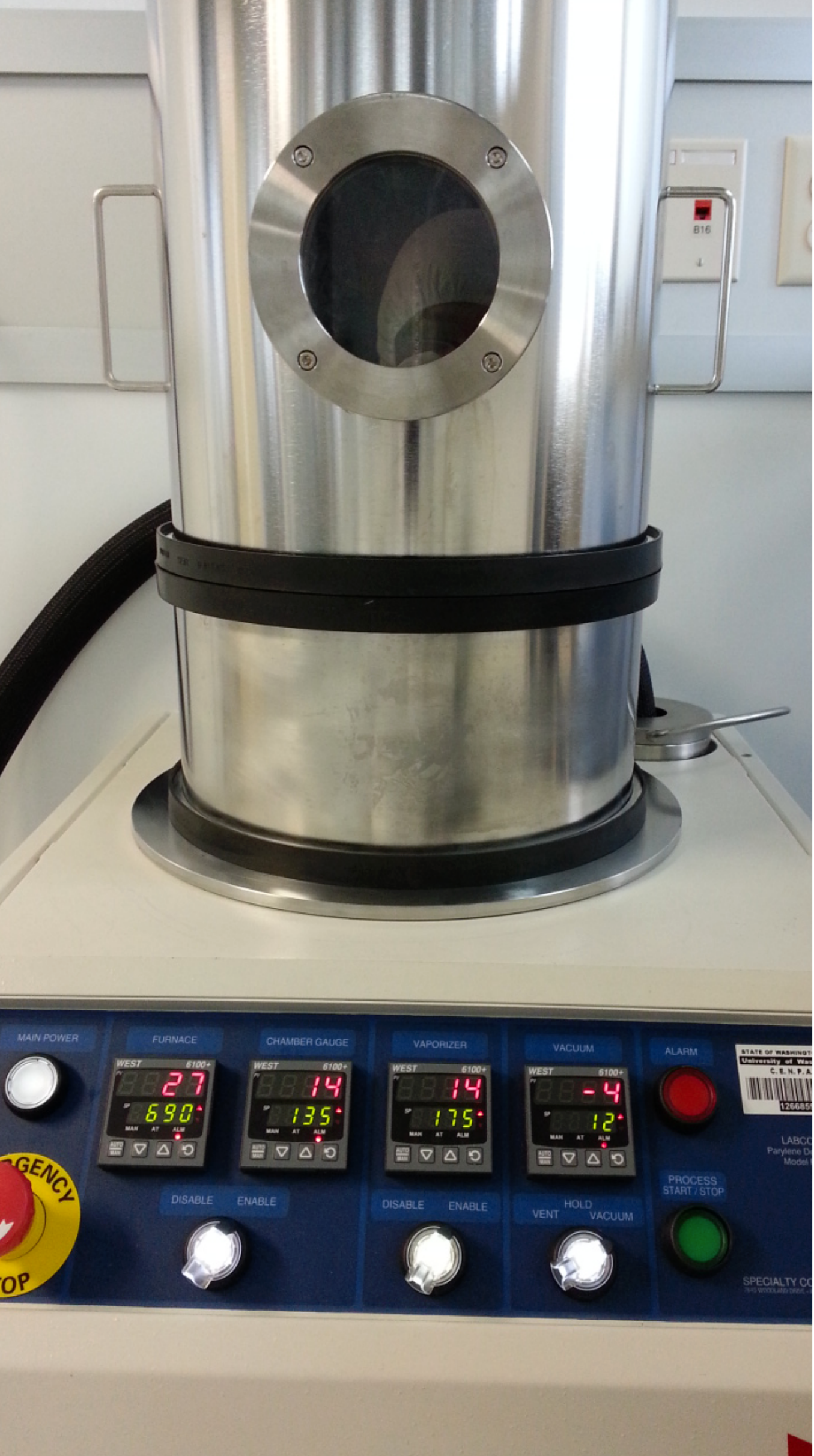}
	\includegraphics[height=6cm, width=4cm]{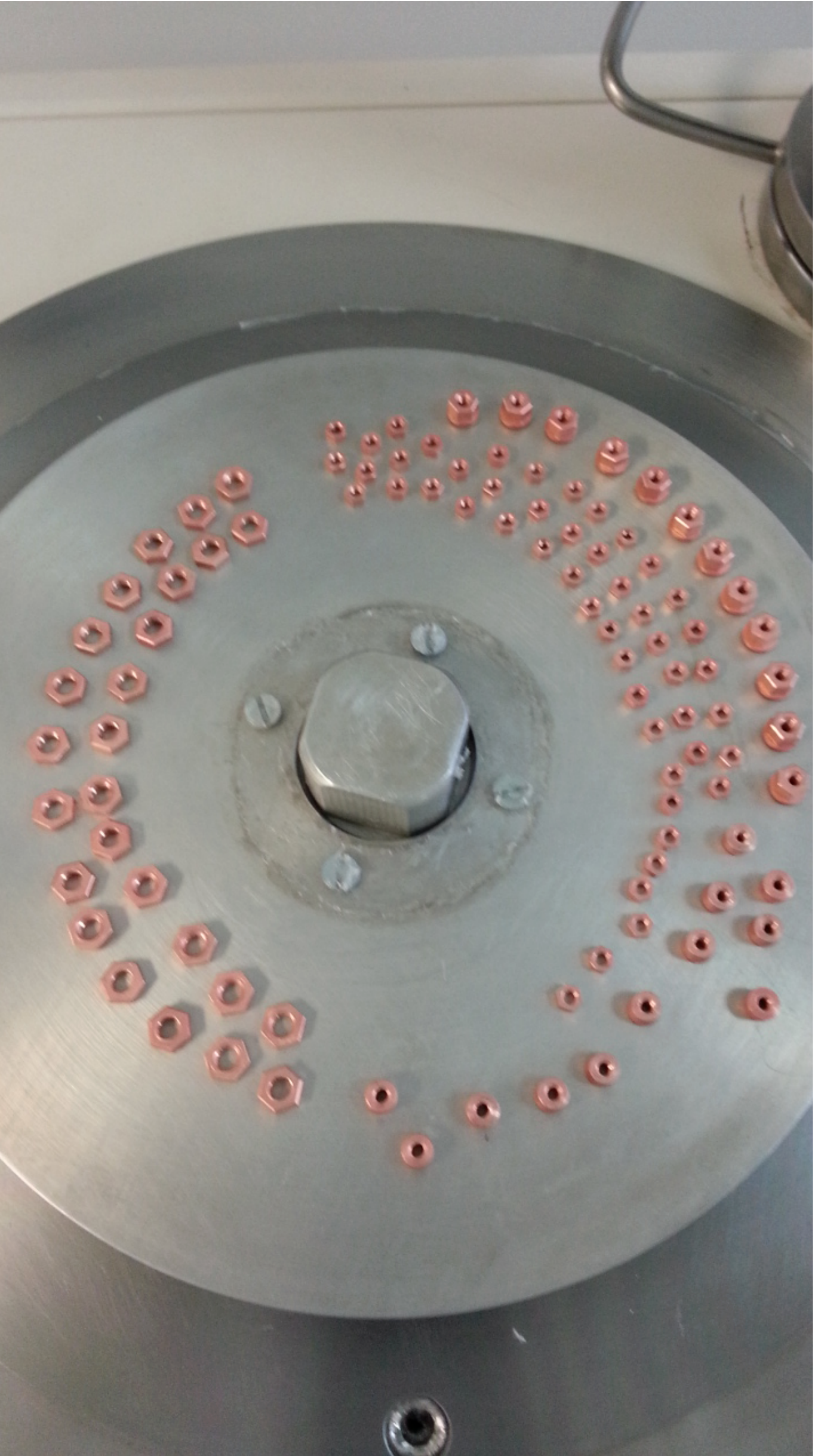}
	\caption{Parylene coating of copper nuts. Left: the parylene coater. Right: parts on the deposition stand.}
	\label{fig:coating}
\end{figure}

Parts were coated with parylene using the PDS 2010 LABCOTER\Rmark\ 2 Parylene Deposition System which was manufactured by Specialty Coating Systems\Tmark\ in Indianapolis, IN. See Figure \ref{fig:coating}. The deposition chamber was operated in a cleanroom located in a surface laboratory at SURF. Because of the high elevation ($\sim$1 mile) on the surface, parts were coated as quickly as possible and returned underground to minimize cosmic ray exposure. Prior to coating, all surfaces of the deposition chamber were wiped with a 2\% solution of Micro-90\Rmark\ that acts as a release agent to ease later cleanup.

The results of our parylene assays are listed in Table~\ref{tab:assay} in Rows~\ref{row:r31} through Row~\ref{row:r44}.  Mirco-90\Rmark\ release agent is listed in Row~\ref{row:r86}. We also assayed parylene film samples from three locations within the deposition chamber as indicated in Rows~\ref{row:r42} through \ref{row:r44}. The observed activities were already low enough for \MJ\ \DEM\ components so further investigation of possible sources of contamination was not pursued.

\subsection{Signal and High-Voltage Cables}
The signal and the high voltage cables used inside the \MJ\ \DEM\ cryostat have very stringent radiopurity requirements as they are in close proximity to the Ge detectors.  For each detector, four coaxial signal cables are needed for connections to the Low Mass Front End (LMFE), and one coaxial cable is required for carrying high voltage to the n$^{+}$ contact.  Axon' Cable~SAS was contracted to manufacture custom miniature Picocoax\Rmark\ cables for these purposes.  The inner dielectric and the outer jacket of the signal (Axon' part number TD11153A) and the high-voltage (Axon' part number TD11153B) cables were extruded from two different stocks of fluorinated ethylene propylene (FEP) --- DuPont\Tmark\ FEP TE9494 and Daikin Neoflon\Tmark\ NP-20.  California Fine Wire Company extruded the central AWG34.5 and AWG40.5 copper conductors in the cables from the same copper stock (CFW heat number 31465), while Axon' Cable extruded the AWG50 copper wires (Axon' batch number MC5092) that form the helical ground shield of the cables.  Assays of the initial cable production runs did not meet the radiopurity requirements.  Several manufacturing steps were identified that were adding unwanted contaminants.  These steps were modified and cleanliness controls were implemented throughout the manufacturing process that resulted in an acceptably radiopure product (Table~\ref{tab:assay}, Rows~\ref{row:r70} through \ref{row:r73}.) The details of these processes are proprietary.

All the raw materials were individually assayed by ICPMS at PNNL for their radiopurity prior to cable production.  After each production step, the partially complete product was washed in solvent in an ultrasonic bath and then baked in a clean oven located in a cleanroom.  Witness samples were collected from each of these steps as quality control.  The unfinished products were stored in clean nylon bags in a cleanroom when not being processed.  The finished cables were wiped with clean isopropyl alcohol, and placed in sealed nylon bags that were filled with dry nitrogen prior to delivery by Axon'.   The cables were stored in an ambient dry nitrogen environment subsequent to their receipt by the \MJ\ collaboration.  

Segments from different parts of the finished cables were assayed by ICPMS.  One spool of each cable, enough for the wirings in the \MJ\ \DEM, was also individually assayed by $\gamma$-ray counting. The resulting assays are listed in Table~\ref{tab:assay} in Rows~\ref{row:r52} through \ref{row:r73}. Parylene coated wire and Mini Coax (Rows~\ref{row:r60} through \ref{row:r65} in Table~\ref{tab:assay}) were not used for the \DEM. The Cu wire used for the parylene coated wire was also from California Fine Wire (Row~\ref{row:r6} in Table~\ref{tab:assay}).

\subsection{Connectors }
In order to complete installation of the \mj\ \dem\ detectors in a clean nitrogen-purged environment to avoid exposure of the germanium crystals to trace radon levels in laboratory air, connections for the four coaxial signal cables must be made within the cryostat. A low-mass, low-radioactivity connector was needed to connect each LMFE's cables to a signal cable bundle that extends outside the shielded space. 

Commercially available connectors were considered, but all options investigated at the time used beryllium-copper (BeCu) contact springs. Measurements of BeCu available in the literature give U and Th specific activity levels on the order of Bq/kg~\cite{Leonard2008,BeCuEdelweiss} and do not satisfy the \mj\ \dem\ radiopurity requirements. Connector pin receptacles intended for use in the \mj\ \dem\ containing miniature BeCu contacts were assayed via ICPMS at Validation Resources, Inc. and were found to have orders-of-magnitude higher specific activity than the same pin receptacles without the BeCu contacts (Table~\ref{tab:assay}, Rows~\ref{row:r90} and \ref{row:r91}). As seen in Table~\ref{PlugBGTable}, these parts alone, if used, would exceed the entire \mj\ \dem\ background budget. 

\begin{figure}
  \centering
     \includegraphics[width=4.5cm]{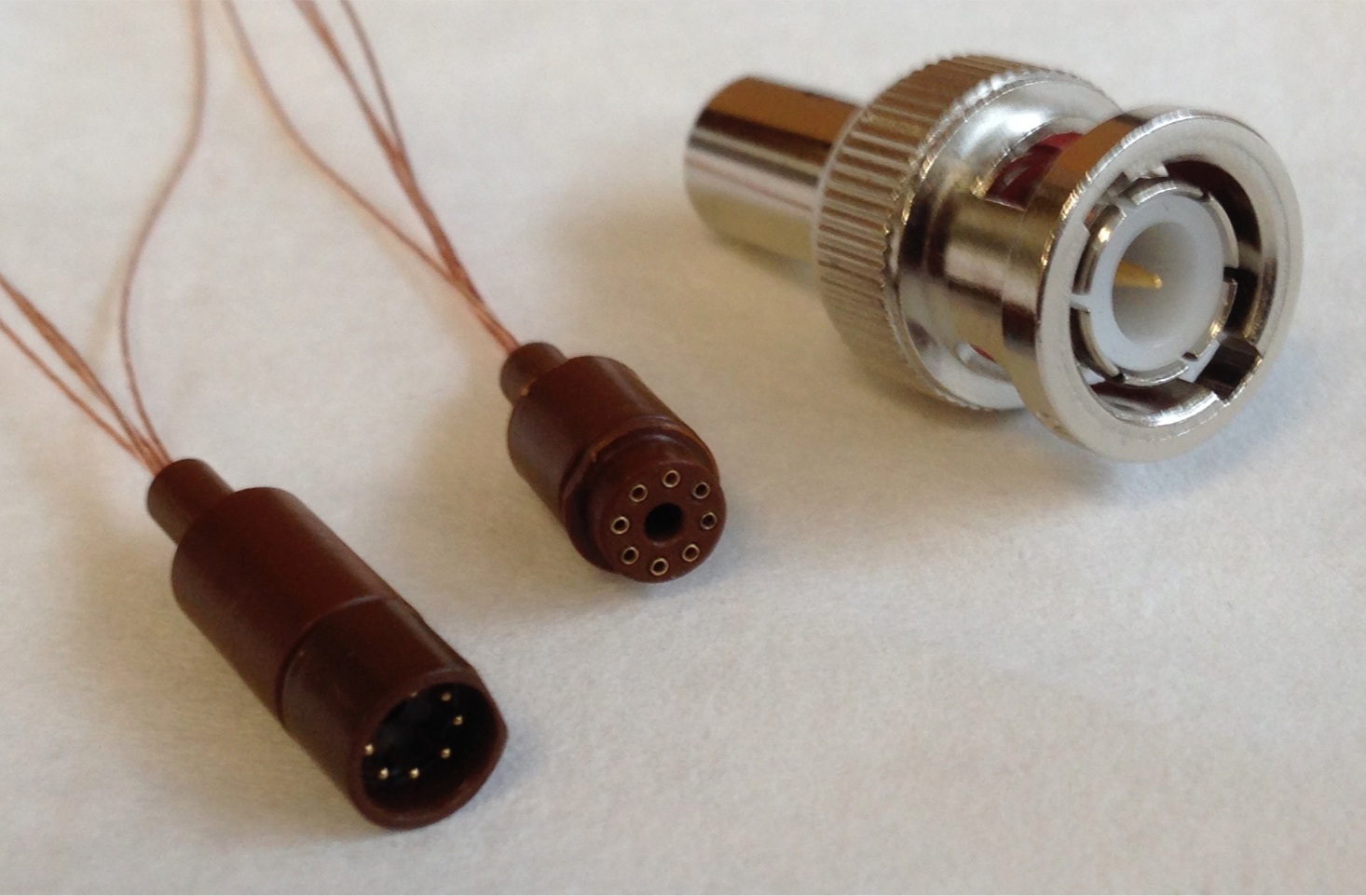}
      \includegraphics[width=5cm,trim= 0 4cm 0 5cm, clip=true]{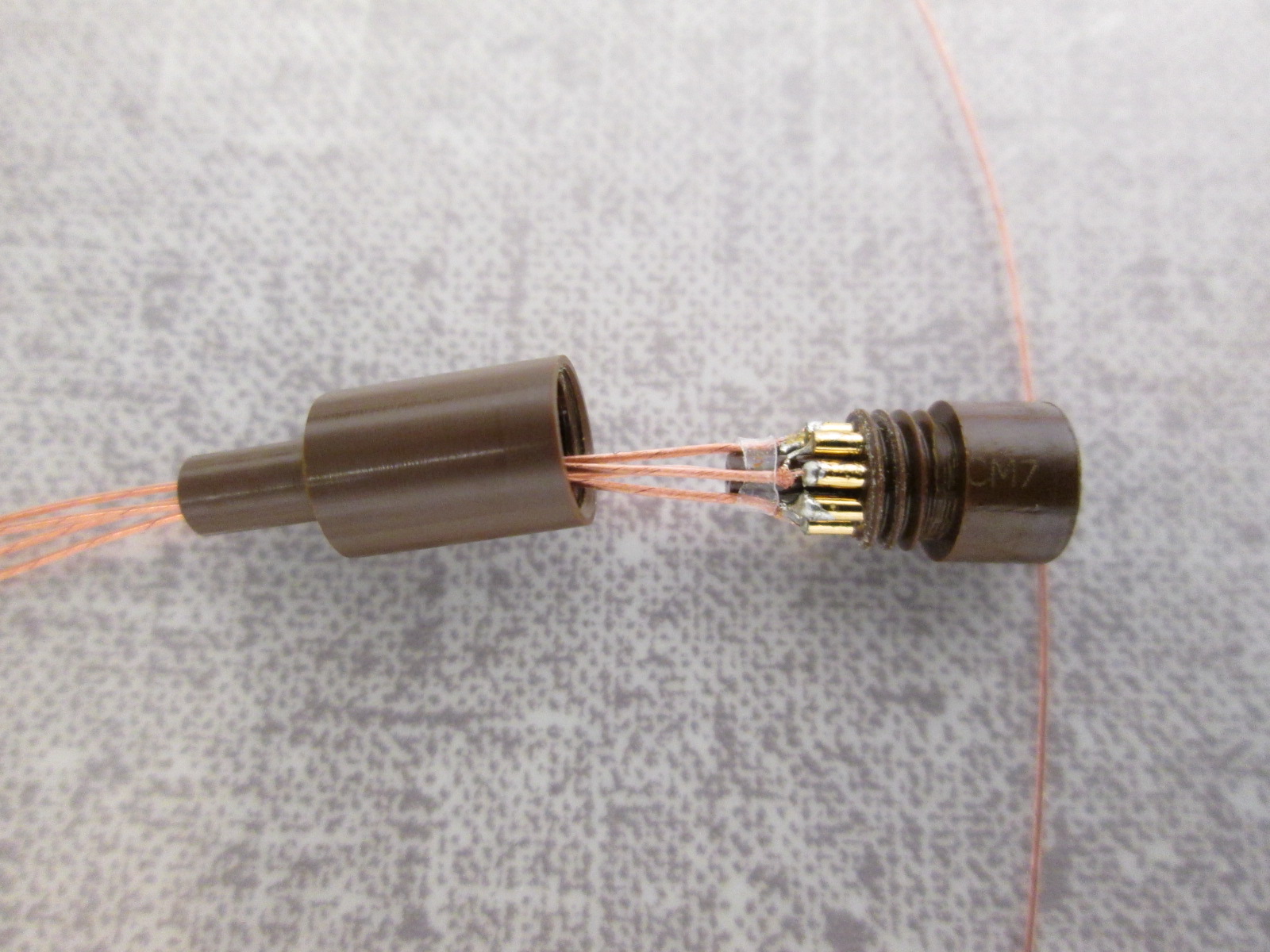}
     \caption{A low-radioactivity signal connector designed for the \mj\ \dem. Mill-Max\Rmark\ pin receptacles are encased in a Vespel\Rmark\ housing, clean-soldered, and strain-relieved with clean FEP heat shrink tubing. Left: \mj\ \dem\ 8-pin signal connector juxtaposed with a male BNC connector. Right: Male plug with end cap removed to show soldering joints and strain relief}
     \label{Both_plugs}
\end{figure}

A connector, shown in Fig.~\ref{Both_plugs}, was designed that uses custom Mill-Max\Rmark\ gold-plated brass pin receptacles without BeCu contact springs. The female socket is Mill-Max\Rmark\ pin receptacle 8210-0-00-15-00-00-03-0, while the male pin uses the post of Mill-Max\Rmark\ pin receptacle 0461-2-00-15-00-00-03-0, whose outer diameter matches the inner diameter of the female socket.  The pins and sockets are slightly misaligned radially, forcing the pin to flex; the restoring force of the pin takes the role of a contact spring in maintaining the electrical connection. The pins and sockets are held in Vespel\Rmark\ SP-1 (polyimide) housings, which are precision-machined at the underground clean machine shop at SURF. The pins, sockets, and Vespel\Rmark\ housings were ultrasonically cleaned in 1\% Micro-90\Rmark\ soap solution,  $>$18-M$\Omega$-cm deionized water, and then Optima\Tmark\ grade ethanol. The Vespel\Rmark\  parts were subsequently leached in 6M Optima\Tmark\ grade nitric acid solution for 72 hours, and then soaked for 24 hours in dionized water to remove absorbed nitric acid. They were then pumped and baked in a vacuum chamber at 100\textdegree C to outgas any remaining liquid, particularly nitric acid which can corrode the brass components.

Following this treatment, the Vespel\Rmark\  parts were stored in a nitrogen-purged environment to prevent radon adsorption. The Axon' coaxial cables were soldered to the connector pins and sockets with a 96.5:3.5 eutectic SnAg alloy solder used in the construction of the Sudbury Neutrino Observatory (SNO) Neutral Current Detectors (NCDs)~\cite{Amsbaugh2007}. 
Purified abietic acid (Rows~\ref{row:r83} and \ref{row:r84} in Table ~\ref{tab:assay}) dissolved in isopropyl alcohol (Row~\ref{row:r115}) was used for flux. This process was performed in a cleanroom under a snorkel to capture soldering fumes. Fluorinated Ethylene-Propylene (FEP) heat-shrink tubing (Rows~\ref{row:r54} through \ref{row:r301}) provided strain relief for the solder joints. To limit exposure of the LMFE components to laboratory air and solder fumes, the LMFE was attached to the cables after soldering was complete.

Individual components were assayed to estimate backgrounds and the results are shown in Table~\ref{PlugBGTable}. Assays of the Vespel\Rmark\  and FEP were performed using NAA (Section~\ref{Sec:NAA}). Assay of pin receptacles, with and without BeCu contact springs and with parts as-is from the manufacturer (no additional cleaning), was performed by Validation Resource, Inc in Bend, OR. (Section~\ref{Sec:ICP-MS}). The results are given in Table~\ref{tab:assay} in Rows~\ref{row:r90} and \ref{row:r91}. Particular attention was given to assaying the solder and related processes. The SnAg eutectic solder from the SNO experiment was re-assayed by the NRCC using GDMS (Row~\ref{row:r400} in Table~\ref{tab:assay}) along with another SnAg eutectic solder newly acquired from Indium Corporation\Rmark listed in Row~\ref{row:r500}. Our result for the NCD solder is very similar to the previous measurement. The solder was melted into a rod shape along with the pure abietic acid and isopropyl alcohol flux before being assayed.  Separate solder samples were prepared using both a new soldering iron tip and a tip that was scraped with a brass scraping pad and dipped in DeoxIT\Rmark\ Tip Tinner \& Cleaner, a mixture of tin and ammonium phosphate used to prevent oxidation of the soldering iron tip. All samples yielded upper limits of $<$0.1-0.3~ppb for both U and Th. There was no significant difference observed between the samples, including the ones prepared with the soldering iron tip that was scraped and dipped in the tip cleaner. The SNO solder is being used to assemble the \mj\ \dem\ signal bundles. 

\begin{center}
\begin{table}[htbp]
\caption{Summary of contaminations and background contributions for signal connector materials. Component masses reported are for one male-female pair. The column labeled Background gives the \MJ\ \DEM\ anticipated background level in the \BBz\ ROI for the measured impurity level assuming the decay chains are in equilibrium. We assume one connector pair for each of 60 detectors in the {\sc Demonstrator}. The row labeled Total indicates the specific activity expected for a connector pair based on the assays and masses of the individual components.  Mass estimates for the solder, solder flux, and shrink tube are uncertain but conservative} 
   \centering
   \begin{tabular}{@{} cccccc @{}} 
      \hline
      \multirow{2}*{Material} & Assay  & Mass         & \multirow{2}*{Isotope	} & Specific Activity     & Background \\
      			      & Method & [g] &                          & [$\mu$Bq/kg] & [\cpRty]	\\
      \hline
      \hline
      \multirow{2}*{Pins (w/BeCu)} 	& \multirow{2}*{ICPMS}	& \multirow{2}*{0.112}	& $^{238}$U & $795000 \pm 12000$ & $8.8 \pm  0.1$ \\
      				   	&			&			& $^{232}$Th& $41000 \pm 1000$	& $2.3 \pm 0.1$ \\
      \hline
      \multirow{2}*{Pins (no BeCu)}	& \multirow{2}*{ICPMS}	& \multirow{2}*{0.112}	& $^{238}$U & $4600 \pm 1500$ 	& $0.05 \pm 0.02$ \\
      					&			&			& $^{232}$Th& $5800 \pm 100$  	& $0.32 \pm 0.01$ \\
      \hline
      \multirow{2}*{Vespel\Rmark\ SP-1}	& \multirow{2}*{NAA}	& \multirow{2}*{0.95}	& $^{238}$U & $<$1000		& $<$0.20 \\
					&			&			& $^{232}$Th& $<$12		& $<$0.01 \\
      \hline
      \multirow{2}*{SnAg Solder}		& \multirow{2}*{GDMS}	& \multirow{2}*{0.04}	& $^{238}$U & $5600 \pm 1000$	& $0.02 \pm 0.004$ \\
					&			&			& $^{232}$Th& $<$12		& $<$0.0002 \\
      \hline
      \multirow{2}*{Solder flux}	& \multirow{2}*{GDMS}	& \multirow{2}*{0.04}	& $^{238}$U & $1200 \pm 200$ 	& $0.005 \pm 0.001$ \\
					&			&			& $^{232}$Th& $<$400		& $<$0.007  \\
      \hline
      \multirow{2}*{shrink tube}	& \multirow{2}*{NAA}	& \multirow{2}*{0.01}	& $^{238}$U & $<$1250		& $<$0.012 \\
      					&			&			& $^{232}$Th& $<$138		& $<$0.007 \\
      \hline
      \multirow{2}*{Total}		&			& \multirow{2}*{1.5}	& $^{238}$U & $<$1500		& $<$0.28 \\
      					&			&			& $^{232}$Th& $<$600		& $<$0.36 \\
      \hline
      \hline
   \end{tabular}
   \label{PlugBGTable}
\end{table}
\end{center}

One each of the completed male and female connectors underwent full-body ICPMS assay to evaluate the cleanliness of the assembly process. The cables were cut from the point at which they enter the end cap housing, and the connectors were transferred to validated quartz crucibles where they were spiked with $^{229}$Th and $^{233}$U tracers. The samples were heated to 700\textdegree C for four hours to ash away polymeric materials, and then digested in a combination of Optima\Tmark\ grade nitric acid, hydrochloric acid, and hydrogen peroxide solutions as described in Ref.~\cite{Overman2013}. A small amount of material was not dissolved (see Table~\ref{PlugFullBodyBGTable}), most likely due to the formation of an insoluble metal oxide precipitate. It is unlikely that the remaining solids contained significant quantities of the analytes, as all solids were dissolved in solution at some point during processing. Since the solids formed during the digestion process, the tracer isotopes account for any analyte adsorption or incorporation into the precipitate. 

After the dissolved samples and precipitate materials were brought to dryness and reconstituted in 8 M nitric acid, the analytes were separated from the sample matrix solution using anion exchange resin. ICPMS was performed as described in Section \ref{Sec:ICP-MS} following the procedure outlined in Ref.~\cite{LaFerriere2015}. The results are given in Table~\ref{PlugFullBodyBGTable} and in Rows~\ref{row:r92} through \ref{row:r94} in Table~\ref{tab:assay}. The process blank for the female plug was lost during sample preparation, but given similar digestion steps and handling in the cleanroom, the process blank generated from the analysis of the male plug provided a good estimate of the levels that would have been seen in the female plug blank. The results are listed in Table~\ref{tab:assay} in Rows~\ref{row:r93} and \ref{row:r94}. Other results related to the connectors are listed in Rows~\ref{row:r83}, ~\ref{row:r84}, ~\ref{row:r86}, ~\ref{row:r74}, and ~\ref{row:r90} through ~\ref{row:r94} in Table~\ref{tab:assay}.

\begin{center}
\begin{table}[htbp]
\caption{Summary of specific activities and corresponding background contributions from full-body ICPMS of two manufactured plugs, one male and one female. The male plug has a larger mass because it uses larger brass pins. The column labeled Background gives the \MJ\ \DEM\ anticipated background level in the \BBz\ ROI for the measured impurity level.} 
\small
   \centering
   \begin{tabular}{@{} ccccccc @{}} 
      \hline
      \multirow{2}*{Material}	& Mass	& Post-Ash	& Undigested	& \multirow{2}*{Isotope} & Specific Activity     	& Background \\[-3pt] 
					& (g)		& Mass (g)	& Mass (g)	&  & ($\mu$Bq/kg)				& (\cpRty) \\
      \hline
      Female 	& \multirow{2}*{0.453}	& \multirow{2}*{0.068} & \multirow{2}*{0.004} & $^{238}$U & $1160\pm17$		& 0.052 \\[-3pt]
      	plug				&		& & 	& $^{232}$Th & $365\pm6$		& 0.083 \\
      \hline
      Male  & \multirow{2}*{0.600}	& \multirow{2}*{0.183} & \multirow{2}*{0.035} & $^{238}$U & $281\pm24$		& 0.017 \\[-3pt]
      	plug				&		& & 	& $^{232}$Th& $27\pm2$			& 0.008 \\
      \hline
      Connector 		& \multirow{2}*{1.05}	& & 	& $^{238}$U & $1160\pm17$		& 0.110 \\[-3pt]
	totals				&			&	&		& $^{232}$Th& $365\pm6$		& 0.174 \\
      \hline
   \end{tabular}
   \label{PlugFullBodyBGTable}
\end{table}
\end{center}

\subsection{The Low-Mass Front End Electronic Board}
\begin{figure}[htdp]
  \centering
    \includegraphics[width = 0.8\textwidth]{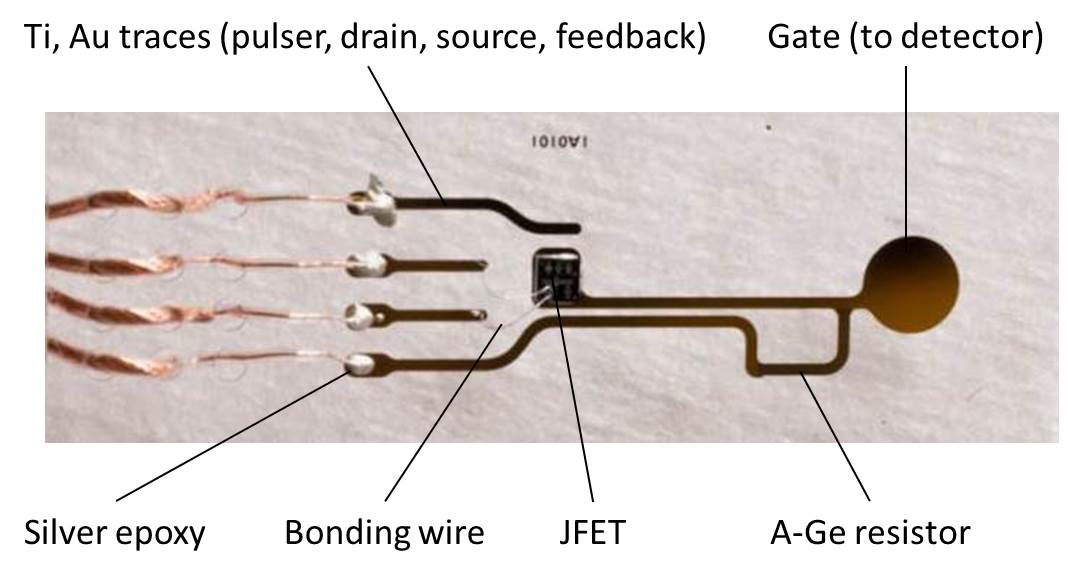}
    \caption{A photograph of an LMFE with wires and JFET attached. The cables were not part of the LMFE assay measurements.}
    \label{fig:LMFE}
\end{figure}

The readout electronics for the \MJ\ \DEM\ were developed with the twin goals of minimizing electronic noise and minimizing radioactivity. In pursuit of the former, the primary stage of amplification was placed close to the output of each detector. In the pursuit of the latter, everything but the primary stage was located outside of the cryostat and those components inside were made as small and as radiopure as possible. The primary amplification stage, the LMFE, is shown in Fig.~\ref{fig:LMFE}~\cite{Barton2011}. It comprises a fused silica substrate on which a photolithographic pattern was used to form conductive traces and a resistor. A JFET was affixed to a gate pad using silver epoxy, and to the drain and source traces. Ultrasonically-drilled holes provide strain relief for the cables, which were bonded to the traces using silver epoxy. The LMFE has a surface area of 145 mm$^2$ and a mass of 80 mg.

The task of controlling the radiopurity of the LMFE presented a number of challenges. The important ones arose from the small size of the LMFE and the novelty of its design. Specifically, determining the level of radioactive contamination in tiny components and controlling exposure to radioactive contamination during a complex production process were key issues. ICPMS was used to measure the radiopurity of all board component materials and the final boards themselves.  ICPMS analysis of all individual components of the LMFE and of three full boards were performed at LBNL, PNNL and Chernogolovka as described in Section~\ref{Sec:ICPMS}. 

Table~\ref{tab:LMFEresults} shows the relative mass fractions of each component of the LMFE board. The substrate accounts for the overwhelming majority of the LMFE mass. It was made from fused silica, a synthetic glass available in extremely high purity due to its use in laser applications. The fused silica was procured in 2-inch diameter, 0.25-mm thick, highly polished wafers, ultrasonically drilled and then thoroughly cleaned before photolithography. Ultrasonic cleaning in  $>$18-M$\Omega$-cm deionized water and 10\% nitric acid was demonstrated to decrease levels of U and Th in the wafers. A number of brands from a variety of manufacturers were tested and those produced by Corning Inc.~were found to be the most promising. Measurements indicated that levels of U and Th decreased with an increase in material grade, cost, and with the distance from the bottom of the manufacturer's original boule. Cost considerations prevented the use of the highest grade material or controlling the position of the material in the boule but, nevertheless, a batch of Corning HPFS\Rmark~7980 standard grade fused silica was obtained with adequate purity. See Rows~\ref{row:r88} through \ref{row:r129}  in Table~\ref{tab:assay}.

The electrical traces on the board and the resistor were sputtered and photolithographically patterned. The traces were 400 nm of gold on top of 20 nm of titanium acting as a bonding layer. The amorphous germanium resistor was 400 nm thick. These films were thin enough that direct ICPMS was impractical. Sufficient masses could only be acquired from longer, dedicated runs in which films $\sim$1-10 microns thick could be deposited. The substrates for these depositions were the same fused silica wafers used for LMFE production and they were initially coated with Micro-90\Rmark\ soap solution as a release agent that allowed the deposited films to be floated from the wafer by slowly submerging them in water. The Micro-90\Rmark\ soap had been assayed to ensure adequate radiopurity (Row~\ref{row:r86}  in Table~\ref{tab:assay}). This procedure was effective for gold and titanium but could not be used for thick films of germanium because they readily lose their integrity with a tendency to flake. For germanium, the metal deposited onto wafers was etched, rather than floated, from the surface. The acids used during etching did not affect the fused silica substrates and the masses of metal removed were determined by before- and after- measurements of the wafer mass. Film-production assay results are listed in Rows~\ref{row:r23} through \ref{row:r605} of Table~\ref{tab:assay}.

Silver epoxy (TRACON TRA-DUCT BA2902) was counted using both ICPMS and $\gamma$-ray counting, with results from both of these methods found to be consistent. Due to its short shelf-life multiple batches of epoxy had to be purchased and counted during the construction of the experiment. The results of samples from the various batches are presented in Table~\ref{tab:assay}.  All batches have been found to be sufficiently pure though significant variations in purity have been observed. Epoxy results from several batches are listed in Rows~\ref{row:r75} through~\ref{row:r85} in Table~\ref{tab:assay}. Part 2 of the epoxy is the hardener. We counted some epoxy in its packing because of its short shelf life and because we wanted to use assayed epoxy.

Bonding wire (99\% aluminum, 1\% silicon manufactured by California Fine Wire) was counted using ICPMS (Row~\ref{row:r89} of Table~\ref{tab:assay}). The cost to acquire enough JFET dies that are used in the LMFE (Moxtek Inc. MX-11) for ICPMS assay was not justified. A substantial batch of dysfunctional MX-17A JFET dies, which have a similar construct as the MX-11 JFET dies, was acquired and assayed using ICPMS (Table~\ref{tab:assay}, Row~\ref{row:r96}). Because it was difficult to obtain sufficient mass of either of these components, the measurements presented are only relatively weak limits.

The second stage of the assay program was the assay of fabricated LMFEs using ICPMS. This was done as a way of accounting for contamination arising during the manufacture process and by contact of the boards with cutting devices or the variety of chemicals during photolithography. Three boards have been counted: one board (LMFE1, Row~\ref{row:r601} in Table~\ref{tab:assay}) consisting of all components except the silver epoxy, JFET and bonding wires; and two complete boards (LMFE2a/b, Rows~\ref{row:r97} and~\ref{row:r98} in Table~\ref{tab:assay}). Boards were dissolved in a microwave reaction system using an ultra high-purity solution of hydroflouric and nitric acids. The use of hydrochloric acid was avoided to eliminate the isobaric interference caused by the molecular ion \nuc{197}{Au}\nuc{35}{Cl}$^+$. Results are given in Table~\ref{tab:LMFEresults} and indicate additional contamination from the production process over what was expected from the assays of the raw materials.

\begin{table}[htdp]
\addtolength{\tabcolsep}{-2pt}
\caption{Summary of purities and activities for LMFE components and manufactured LMFE boards. The row labeled Total is the predicted activity of a completed LMFE based on component activity and mass. Activities of completed boards are elevated compared to what one would expect from the individual components. The excess is assumed to arise from contamination during the assembly process and assembly was moved to a high-quality clean room after these results were obtained.  In the first column the numbers in parenthesis refer to rows in Table~\ref{tab:assay}.}
\begin{center}
\begin{tabular}{lrrrrrrr}
\hline

\mcl{1}{Item}  & \mcc{1}{Mass} & & \mcc{2}{Concentration (ppb)} & & \mcc{2}{Activity (nBq)} \\

\cline{4-5} \cline{7-8} 

          & \mcc{1}{fraction (\%)} & & \mcc{1}{\nuc{232}{Th}} & \mcc{1}{\nuc{238}{U}} & & \mcc{1}{\nuc{232}{Th}} & \mcc{1}{\nuc{238}{U}} \\
         
\hline

Fused silica (\ref{row:r602})	&  97.24\phantom{00}                   & &    $<$ 0.01   & 0.10$\pm$0.01 & & $<$ 3.2           & 99$\pm$10       \\
Amorph. Ge (\ref{row:r603})        &  0.0029  & &   2.4$\pm$0.7  & 1.7$\pm$0.4     & & 0.02$\pm$0.01   & 0.05$\pm$0.01 \\
Au traces (\ref{row:r604})  		    &  0.114\phantom{0}  & &      $47\pm1$ & 2.0$\pm$0.3     & &  17.9$\pm$0.4          & 2.3$\pm$0.3 \\
Ti traces (\ref{row:r605})		    &  0.0013& &     $<$ 0.4    &	$<$ 0.1        & & $<$ 0.002      & $<$ 0.001 \\
JFET die (\ref{row:r96})		       &  \phantom{0}1.17\phantom{000}                                & &	     $<$ 2.0  &   $<$ 0.14     & & $<$ 7.8           & $<$ 1.7 \\
Al-Si wire (\ref{row:r89})       	&  0.0049  & &	     91$\pm$2  &  8.7$\pm$0.4    & & $<$ 1.50$\pm$0.03 & 0.44$\pm$0.02 \\
Silver epoxy (\ref{row:r81})	 		& \phantom{0}1.46\phantom{00}                                 & &         $<$5   &  20$\pm$3        & & $<$ 24             & 300$\pm$45  \\
				
\hline
										
Total	 		            &   100.\phantom{0000}             & &               &	                & &	$<$ $\sim$ 55	&	$<$ $\sim$ 402	\\

\hline
										
LMFE1 (\ref{row:r601})	          &                                           & &  0.13$\pm$0.01 &  0.43$\pm$0.04 &	&         42$\pm$3  &	44$\pm$4	\\
LMFE2a (\ref{row:r97})	      &                                           & &  2.12$\pm$0.06  &   0.85$\pm$0.03 &	&       706$\pm$2  &	87$\pm$3	\\
LMFE2b (\ref{row:r98})	      &                                           & &  1.61$\pm$0.03  &   0.85$\pm$0.03 &	&     536$\pm$10  &	87$\pm$3	\\
\hline

\end{tabular}
\end{center}
\label{tab:LMFEresults}
\end{table}

\subsection{TIG Welding}
Early designs of the \DEM\ considered using tungsten inert gas welding for certain copper joints. Although in the end, this technique was not used, we include our study of its potential for radioactive contamination for completeness.
	A comparative study was made of the characteristics of zirconiated tungsten cathodes (W+1 wt\% ZrO$_2$) and ceriated tungsten electrodes (W+2 wt\% CeO$_2$).  Eight electrodes with a radius of 0.12 cm, and lengths varying from 3.33 to 5.7 cm were used.
	The TIG(Zr) rods (26.5 g) were $\gamma$-ray counted for 57 days. The $\gamma$-ray count for the TIG(Ce) rods (25.3 g) was only 0.64 days due to the high activity.  Most of the activity appears to be Th contamination, indicating that these rods contain approximately 0.1\% ThO$_2$.  The assay results are listed in Table~\ref{tab:assay} in Rows~\ref{row:r25} and \ref{row:r26}.
	
Once the assay results were completed, erosion measurements were made of the two different types of cylindrical cathodes.  The cathodes were used on a dc-transferred torch operating at atmospheric pressure.  New gas cups and lenses were used on the torch body.  An ultra-high purity He (99\%) gas was used as the shielding gas.  Erosion measurements were taken with each cathode at current intensities ranging from 100-130 Amps, with an arc voltage of 18-19 V. The results of the erosion tests for the the TIG(Zr) and TIG(Ce) tips show that the mass loss was near the uncertainty in our balance. Using the mass measurements, upper limits on the average erosion rate for the TIG(Zr) and TIG(Ce) tips were determined to be 0.280 mg/cm and 0.568 mg/cm, respectively. The TIG tips were examined and there did not appear to be any spattering from the welds that could result in an increase in weight.   

In order to study the effects of any weld contamination on the \dem, a simulation of a small section of the Cu tubing was performed. This tubing was considered for use as calibration-source access to the \dem. The simulated weld joint was 1.57 cm. Within the weld, we would expect to deposit less than 0.4 mg of the TIG(Zr) electrode, and less than 0.9 mg of the TIG(Ce) electrode.  Based on the simulations, the experimental background was predicted to be less than 0.024 and 6.96 \cpRty\ from the TIG(Zr)  and TIG(Ce) electrode, respectively.   A previous experiment found that TIG(Zr) electrodes consume around 0.00224 g/ft in a weld in Ar gas \cite{win57}.  Using this electrode consumption estimate, 0.0118 mg of the tip would be embedded in the weld resulting in a background level of 0.0008 \cpRty.  

\subsection{Other Assayed Materials}
There are a number of materials in Table~\ref{tab:assay} that were used, or were under consideration for use, in the \DEM\ but were not discussed in the previous subsections. Here we summarize those materials that have not yet been described but reside in the table.

Teflon\Rmark\ (Row~\ref{row:r35}), Peek\Rmark\ (Row~\ref{row:r34}), and Vespel\Rmark\ (Rows~\ref{row:r36} through \ref{row:r38}) were all used within the detector string assembly (in addition to uses mentioned previously).  These samples were assayed via NAA and were all found to be quite pure, although in several cases sensitivity to U was limited by the presence of \nuc{24}{Na} and \nuc{82}{Br}. The highest-sensitivity assay was performed on our Teflon material, which was identified by the EXO collaboration~\cite{Leonard2008} and is used as the primary structural plastic in the innermost regions of the Demonstrator. 

The electroformed Cu pins that provide electrical contact between the LMFE and the Ge crystal were coated with a thin layer of Sn. We assayed two Sn samples for this purpose (Rows~\ref{row:r19} through \ref{row:r21}). $\gamma$-ray counting assayed the entire Sn supply, whereas ICPMS assayed a small sample of the same source material. We do not understand fully why the two assay techniques gave different results in this case. The Sn is used in very small quantities and even at the highest assay value doesn't significantly affect our background model. 

The shield includes a high density polyethylene layer (Row~\ref{row:r45}) and a 5\%-boron-loaded high density polyethylene layer (Row~\ref{row:r46}). It also includes a layer of OFHC copper from Southern Copper (Rows~\ref{row:r3} and \ref{row:r4}).

We investigated several options for alternative cryostat seals. Indium was rejected due to its natural radioactivity, and existing measurements of butyl also indicated it would be too radioactive. We assayed o-rings made of Kalrez using NAA (Rows~\ref{row:r50} and \ref{row:r51}), these were somewhat cleaner but still too radioactive for use in the \DEM. We found that we could make seals of sufficient purity using Parylene film or PTFE (Row~\ref{row:r302}).

The electronic feedthroughs from the vacuum were situated on conflat flanges attached to a 6-way cube intersection (Row~\ref{row:r22}). Because there is a shine path down the cross arm into the shield, this cube was assayed and included in the background model. Shielding plates in the cross arm make us insensitive to the relatively high Th levels in this cube.

The calibration system uses a Teflon\Rmark\ tube (Row~\ref{row:r49}) as a guiding track that extends through the shield and encircles the cryostat. The position of a calibration line source is controlled by a motor controller and drive wheels that insert and remove the source from this track. Because the drive wheels contact the source and therefore abraded material might, in principle, be left within the track, the drive wheels were assayed (Row~\ref{row:r99}).

Sapphire was considered as a material for the LMFE, but was not used due to the progress using fused silica. We assayed sapphire samples using NAA and the results are listed in Row~\ref{row:r95}. The U value was limited by the presence of \nuc{24}{Na} in the activated sample.

Our sources at SURF are checked for leakage on a regular schedule. We assayed the KIMTECH PURE\Rmark\ W4 critical task wipers from Kimberly-Clark Professionals\Rmark\ (Row~\ref{row:r100}) used to swipe the sources to better understand the background to those measurements.

Our nitrogen purge system uses N$_2$ gas purified by passing through a charcoal trap. We also used selected charcoal as a getter in the cryostats used for detector transport. We studied a number of charcoals for their radiopurity (Rows~\ref{row:r101} through \ref{row:r106}).

The purity of glass beads was determined (Rows~\ref{row:r117} through \ref{row:r125}). These beads were considered for use as a cable ID mechanism but was not implemented.

Epoxies were assayed to identify a product to fix the LMFE's onto their electroformed Cu mount. Hysol\Rmark\  0151\Tmark\ and 1C\Tmark\ (Rows~\ref{row:r108} through \ref{row:r111}) were considered and the Hysol\Rmark\ 0151\Tmark\ was chosen. Torr Seal\Rmark\ (Rows~\ref{row:r112} and \ref{row:r113}) was considered for mounting the HV connector socket at the feedthrough flange. Hysol\Rmark\ was used instead.

Silicone Rubber (Row~\ref{row:r114}) was considered for use in potting the in-vacuum-side of HV feedthroughs. This was not implemented.

\section{Background Projections and Conclusion.}
\label{Sec:Conclusions}
The assay program described in this article has provided values for the radioactive contamination of the materials and components that can be used to estimate the background from trace quantities of U and Th in the \DEM\ apparatus. Such estimates were made via simulation of the geometry of the \DEM\ by assigning the measured impurity level to each component that comprises the apparatus. The simulation was based on the MaGe Monte Carlo framework described in Ref.~\cite{Bos11}. We use the estimates to determine how many counts will remain in a fully analyzed spectrum arising from the measured levels. As mentioned in the Introduction, our goal for the \DEM\ is \threecpRty. We quote each contribution from each material in similar units in Table~\ref{tab:BGsummary}. We did identify that contamination arising from machining and handling will require further future research. 

The U and Th impurities in detector components do not represent a full background model for the \DEM. For completeness, we include Table~\ref{tab:allBGsummary}, which groups radiogenic background contributions by detector component. This table also includes background contributions from cosmogenic isotopes produced in materials while they resided on the Earth's surface, $\gamma$ rays originating from external to the apparatus, $\alpha$ particles from Rn daughters plating out on surfaces of the apparatus, neutron interactions, direct passage of cosmic ray $\mu$, and neutrino backgrounds. The quoted values are the results of our  simulations and other calculation methods, which will be detailed in future publications.

The background budget for the \MJ\ \DEM\ indicates that a large experiment built with a tonne of isotope should be feasible with some modest improvements.

\begin{table}[htdp]
\caption{The summary of the contributions to the background based on the assay results given in this article, grouped by detector material. The background values assume that the radioactive chains are in equilibrium. }
\small
\begin{center}
\begin{tabular}{llcrr}
\hline
\hline

\multirow{2}{*}{Material}  & \multirow{2}{*}{Typical Use} &  Decay  & \multicolumn{2}{c}{Achieved Assay} \\

\cline{4-5}  

          & &  Chain & $\mu$Bq/kg & \cpRty \\
         
\hline

\multirow{3}{*}{Electroformed Cu}	& Inner Cu Shield, Cryostat, & Th & \phantom{0}0.06 & \phantom{0}0.15 \\
\cline{3-5}
						& Coldplate, Thermal Shield, & U & \phantom{0}0.17 & \phantom{0}0.08 \\
\cline{3-5}
						& Detector Mounts				&	&			&				\\
\hline
\multirow{2}{*}{OFHC}	&\multirow{2}{*}{Outer Copper Shield}	& Th		& \phantom{0}1.1\phantom{0}	& \phantom{0}0.26 \\
\cline{3-5}		
						&							& U		&\phantom{0}1.25				&\phantom{0}0.03	\\	

\hline		
\multirow{2}{*}{Pb	}	&\multirow{2}{*}{Lead Shield}			& Th		& 5\phantom{00}	& 0.26\\
\cline{3-5}		
						&							& U		& $$36\phantom{00}	&$$0.37	\\	

\hline		
\multirow{2}{*}{PTFE	}	&\multirow{2}{*}{Detector Supports}		& Th		& \phantom{0}0.1\phantom{0}	& \phantom{0}0.01 \\
\cline{3-5}		
						&							& U		& $<$5\phantom{00}	&$<$0.01	\\	

\hline		
\multirow{2}{*}{Vespel	}	&Coldplate Supports,			& Th		& $<$12\phantom{00}	& $<$0.01 \\
\cline{3-5}		
						&	Connectors				& U		& $<$1050\phantom{00}	&$<$0.4\phantom{0}	\\	

\hline		
\multirow{2}{*}{Parylene	}	&Cu Coating,					& Th		& 2150\phantom{00}	& 0.27 \\
\cline{3-5}		
						&	Cryostat Seals				& U		& 3110\phantom{00}	&0.09	\\	

\hline		
	Silica,	&Front-end								& Th		& 6530\phantom{00}	& 0.32 \\
\cline{3-5}		
	Au, Epoxy&	Electronics							& U		& 10570\phantom{00}	&0.28	\\	

\hline		
	Cu Wire,	&Cables		& Th							& 2.2\phantom{0}	& 0.01 \\
\cline{3-5}		
	\& FEP	&			& U							& 145\phantom{00}	&0.08	\\	

\hline		
	Stainless	&\multirow{2}{*}{Service Body}					& Th		& 13000\phantom{00}	& $<$0.04 \\
\cline{3-5}		
	Steel		&										& U		& $<$5000\phantom{00}	&$<$0.03	\\	

\hline		
	Solder	&\multirow{2}{*}{Connectors}					& Th		& 210\phantom{00}	& 0.13 \\
\cline{3-5}		
	Flux		&										& U		& 335\phantom{00}	&0.06	\\	
\hline\hline

\end{tabular}
\end{center}
\label{tab:BGsummary}
\end{table}

\begin{table}[htdp]
\caption{The summary of all the backgrounds contributing to the \DEM, with radiogenic backgrounds grouped by detector component. The background values assume that the radioactive chains are in equilibrium.}
\begin{center}
\begin{tabular}{lr}
\hline
\hline
Background Contribution									& Rate \\
													& \cpRty  \\
\hline
Electroformed Cu										& 	0.23	\\
OFHC Cu Shielding										&	0.29	\\
Pb Shielding											&      0.63	\\
Cables and Internal Connectors							&$<$0.38	\\
Front Ends											&	0.6\phantom{0}	\\
U/Th within the Ge										&$<$0.07	\\
Plastics + Other										&	0.39	\\
\nuc{68}{Ge}, \nuc{60}{Co} within the \nuc{enr}{Ge}				&	0.07	\\
\nuc{60}{Co} within the Cu									&	0.09	\\
External $\gamma$ rays, (alpha,n) Reactions					&	0.1\phantom{0}	\\
Rn and Surface $\alpha$ Emission							&	0.05	\\
Ge, Cu, Pb (n,n'gamma) Reactions							&	0.21\	\\
Ge(n,n')	 Reactions									&	0.17	\\
Ge(n,$\gamma$)										&	0.13	\\
Direct $\mu$ Passage									&	0.03 \\
$\nu$ Induced Background								&$<$0.01	\\
\hline
Total													& $<$3.5\phantom{0}	\\
\hline\hline
\end{tabular}
\end{center}
\label{tab:allBGsummary}
\end{table}

\section*{Acknowledgments}
This material is based upon work supported by the U.S. Department of Energy, Office of Science, Office of Nuclear Physics under Award  Numbers DE-AC02-05CH11231, DE-AC52-06NA25396, DE-FG02-97ER41041, \\DE-FG02-97ER41033, DE-FG02-97ER41042, DE-SC0012612, DE-FG02-10ER41715, DE-SC0010254, and DE-FG02-97ER41020. We acknowledge support from the Particle Astrophysics Program and Nuclear Physics Program of the National Science Foundation through grant numbers PHY-0919270, PHY-1003940, 0855314, PHY-1202950, MRI 0923142 and 1003399. We acknowledge support from the Russian Foundation for Basic Research, grant No. 15-02-02919. We  acknowledge the support of the U.S. Department of Energy through the LANL/LDRD Program.

This research used resources of the Oak Ridge Leadership Computing Facility, which is a DOE Office of Science User Facility supported under Contract DE-AC05-00OR22725. This research used resources of the National Energy Research Scientific Computing Center, a DOE Office of Science User Facility supported under Contract No. DE-AC02-05CH11231. We thank our hosts and colleagues at the Sanford Underground Research Facility for their support. We acknowledge support from the facilities; the Waste Isolation Pilot Plant, the Kimballton Underground Research Facility and the Low Background Facility at Oroville, the Berkeley Low Background Facilities at SURF and at LBNL, the McClellan Nuclear Radiation Center, the North Carolina State University's PULSTAR research and teaching reactor, the Radiation Safety Office at Virginia Tech, the Radiation Safety Department at the University of Tennesse, Knoxville, and Nu Instruments Limited in Wrexham, UK.

\bibliographystyle{iopart-num.bst}
\bibliography{DoubleBetaDecay.bbl}

\end{document}